\documentclass[final,3p,times,twocolumn]{elsarticle}
\biboptions{sort&compress}

\usepackage{amsmath}             
\usepackage{bm}                  
\usepackage{graphicx}

\usepackage[normalem]{ulem} 
\usepackage{units} 
\usepackage{float} 
\usepackage{hyperref}
\hypersetup{
	colorlinks=true,
	linkcolor=blue,
	urlcolor=blue,
	citecolor=blue,
	pdfpagemode=FullScreen,
}

\usepackage{amssymb}
\usepackage{amsfonts}

\usepackage{etoolbox}
\patchcmd{\MaketitleBox}{\footnotesize\itshape\elsaddress\par\vskip36pt}{\footnotesize\itshape\elsaddress\par\parbox[b][36pt]{\linewidth}{\vfill\hfill\textnormal{Draft of \today}\hfill\null\vfill}}{}{}%

\AtBeginDocument{\mathcode`v=\varv} 

\def\nablab{{\bm \nabla}}

\def\figures{.}

\usepackage{color}
\definecolor{gray}{rgb}{0.5,0.5,0.5}
\definecolor{lgray}{rgb}{0.8,0.8,0.8}
\definecolor{dgray}{rgb}{0.6,0.6,0.6}
\definecolor{dred}{rgb}{0.5,0.0,0.0}
\definecolor{dgreen}{rgb}{0.0,0.5,0.0}
\definecolor{dblue}{rgb}{0.0,0.0,0.5}
\definecolor{violet}{rgb}{0.7,0.0,0.5}

\definecolor{lred}{rgb}{1.0,0.5,0.5}
\definecolor{lgreen}{rgb}{0.5,1.0,0.5}
\definecolor{lblue}{rgb}{0.5,0.5,1.0}

\def\A{{\mathcal A}}

\begin{document}
	
\begin{frontmatter}
	\title{\vspace{-1.4cm}Study of low-frequency core-edge coupling in a tokamak: \\ III. Core-localized MHD continuum pulsations \& distant forced reconnection\vspace{-0.5cm}}
	
	\author[naka]{Andreas~Bierwage\corref{cor1}}
	\author[naka]{Panith~Adulsiriswad}
	\author[kaist]{Gyungjin~Choi}
	\author[kaist]{Young-chul~Ghim}
	\author[rokk]{Masatoshi~Yagi\corref{cor2}\vspace{-0.1cm}}
	
	\cortext[cor1]{{\it Email address:} {\tt bierwage.andreas@qst.go.jp}}
	\cortext[cor2]{{\it Present affiliation:} Research Organization for Information Science and Technology (RIST), Tokyo 105-0013, Japan}
	
	\address[naka]{National Institutes for Quantum Science and Technology (QST), Naka Institute for Fusion Science and Technology, Ibaraki 311-0193, Japan}
	\address[kaist]{Department of Nuclear and Quantum Engineering, KAIST, Daejeon 34141, Republic of Korea}
	\address[rokk]{QST, Rokkasho Institute for Fusion Energy, Aomori 039-3212, Japan\vspace{-0.45cm}}

\begin{abstract}
	Slow magnetoacoustic pulsations (SMAPs) are found in MHD simulations of a tokamak plasma whose safety factor $q$ near the center is flat and slightly above unity ($q \gtrsim 1$). SMAPs are located on the central plateau of the slow magnetoacoustic continuum $\omega_{\rm S} = k_\parallel c_{\rm S}$, where $c_{\rm S}$ is the speed of sound and $k_\parallel$ the wavenumber parallel to the magnetic field. In our model, SMAPs exist when the ion viscosity and thermal diffusivity are sufficiently low. They can be driven unstable by a pressure gradient in the $q \sim 1$ region when the electric resistivity is sufficiently high. Exponentially growing SMAPs consist of standing slow waves on magnetic surfaces that are radially synchronized into a quasi-interchange structure with poloidal/toroidal mode numbers $m/n=1/1$. After free energy depletion, saturated quasi-linear pulsations (alternating $n=1$ and $0$) in the central $q\sim 1$ region couple to distant $q\geq 2$ rational surfaces that --- in simulations with high resistivity and correspondingly strong loop voltage --- undergo ``reversible'' magnetic reconnection: as the magnetic islands wax and wane with period $2\pi/\omega_{\rm S}$, their X- and O-points alternate. These results show how the MHD model facilitates non-local coupling of slow waves, pressure-driven resistive interchange and tearing. This motivates further study in kinetic models, where collisionless mechanisms for fast reversible reconnection exist and proper treatment of parallel dynamics will allow to assess the role of Landau damping as well as the question whether the ${\bm B}$ field's weak ergodicity in the $q\sim 1$ region allows the waves to outpace the ion's parallel streaming to maintain the thermal misbalance underlying SMAPs. Also of interest are pulsations closer to the Alfv\'{e}nic branch, satisfying $\omega \approx k_\parallel v_{\rm A}$ with Alfv\'{e}n speed $v_{\rm A}$, which require no resistivity and are less dependent on thermal misbalance.%
\end{abstract} 

\begin{keyword}
	Tokamak plasma \sep Pulsations \sep Slow magnetoacoustic wave \sep Magnetic reconnection \sep Non-local effects\vspace{-0.3cm}
\end{keyword}
\end{frontmatter}

\thispagestyle{empty} 
\pagestyle{plain} 
\everypar{\looseness=-1} 

\renewcommand\contentsname{\vspace{-0.9cm}}
\tableofcontents

\section{Introduction}
\label{sec:intro}

\begin{figure}
	[t!]\vspace{-0.7cm}
	\centering
	\includegraphics[width=0.48\textwidth]{\figures/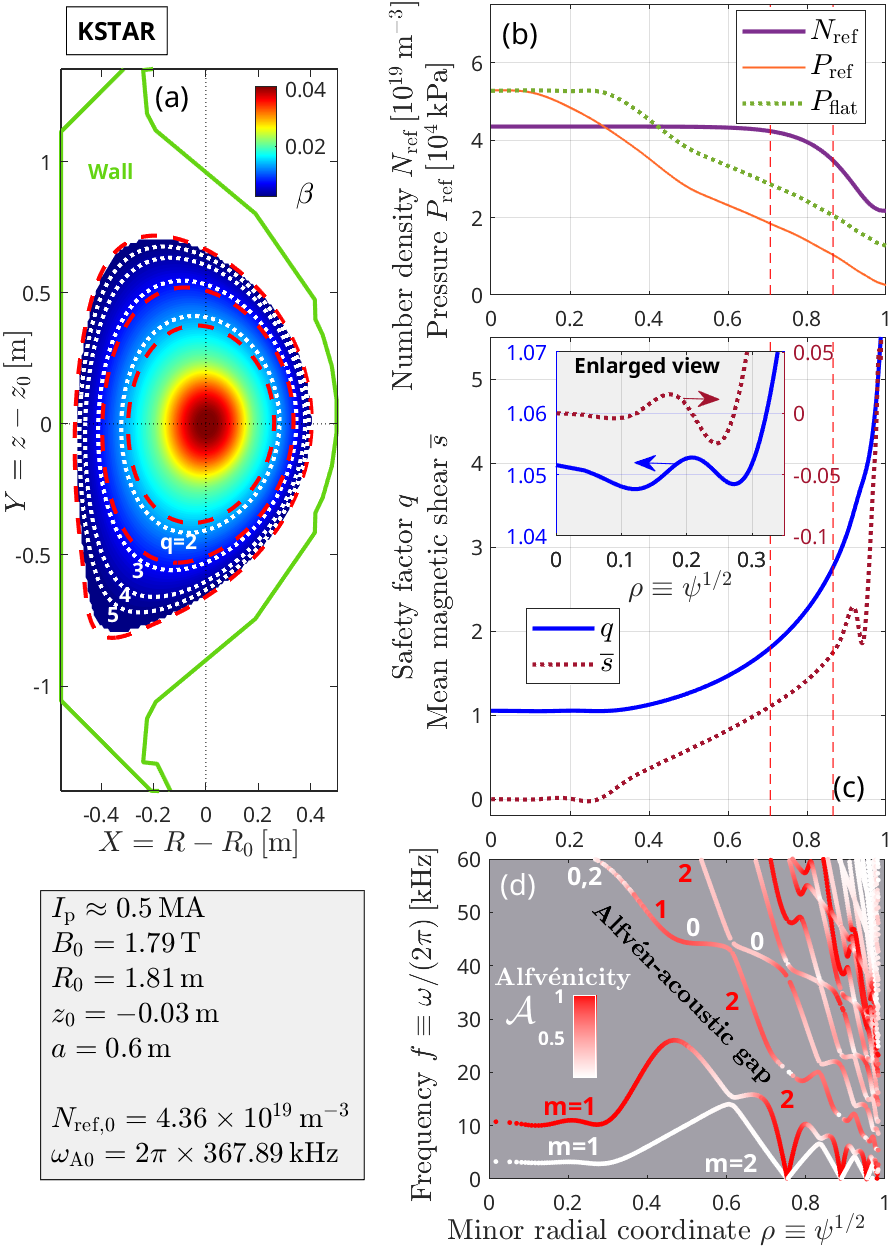}\vspace{-0.1cm}
	\caption{KSTAR-like plasma used in our simulations. This case is similar to ``Model 1'' of Ref.~\protect\cite{Bierwage26a}, except that the density and pressure profiles were modified in preparation for an ongoing benchmarking study. The plasma's poloidal $(R,z)$ cross-section in panel (a) shows the toroidal beta ($\beta$, parula color contours), magnetic flux surfaces with $q = 2,3,4,5$ (white) and the wall (green). Red dashed lines indicate the surfaces corresponding to $50\%$, $75\%$ and $100\%$ of the poloidal flux space, where we applied a non-slip boundary condition in simulations with different domain sizes. The box below panel (a) lists the values of the plasma current $I_{\rm p}$, on-axis magnetic field strength $B_0$, coordinates of the magnetic axis $(R_0,z_0)$, mean minor radius $a$, on-axis number density $N_{\rm ref,0}$, and on-axis Alfv\'{e}n frequency $\omega_{\rm A0} \equiv v_{\rm A0}/R_0$. Panel (b) shows the reference profiles for the number density $N_{\rm ref}(\rho)$ and pressure $P_{\rm ref}(\rho)$ as functions of the square root of the normalized poloidal flux $\rho \equiv \psi^{1/2}$. In addition, we have constructed a centrally flat pressure profile $P_{\rm flat}(\rho)$ (green dotted line) that will allow us to clarify the role of the pressure gradient. Panel (c) shows the profiles of the safety factor $q(\rho)$ and mean magnetic shear $\overline{s}$. This case has a nearly flat plateau with $q \approx 1.05$ and $|\overline{s}| \lesssim 0.02$ near the axis ($\rho \lesssim 0.3$) as shown in the inset. Panel (d) shows the ideal MHD continuous spectra computed by {\tt FALCON} \protect\cite{Falessi20, Falessi19b}. The local Alfv\'{e}nicity $\A$ is indicated by shades of red (Alfv\'{e}nic, $\mathcal{A}\rightarrow 1$) to white (acoustic, $\mathcal{A} \rightarrow 0$).}
\label{fig:01_setup}\vspace{-0.2cm}
\end{figure}

Tokamak plasmas that have wide regions where the safety factor $q$ is flat and close to unity ($q \sim 1$) --- so that the magnetic shear is weak and magnetic field lines nearly (but not quite) close on themselves after one poloidal and toroidal turn --- are interesting both from the fundamental plasma physics and practical points of view. In particular, we are interested in configurations where a significant portion of the core plasma has $q\sim 1$. This is typical for the so-called ``hybrid scenario'' \cite{Gormezano07, Yoshida25}, and Fig.~\ref{fig:01_setup} shows an example based on a KSTAR tokamak plasma \cite{Lee23, Lee26} as modeled in a recent numerical study \cite{Bierwage26a}. From the fusion reactor application point of view, such configurations have attracted attention due to their relevance for self-organized steady-state operation and improved confinement, including experiments \cite{Gruber99, Luce01, Sips02, Isayama03, Oyama09, Kim16, Hobirk23, Ko24, Litaudon26} and DEMO power plant designs \cite{ChenJ21, Sugiyama24, Coleman25}. From the physics side, these configurations are interesting because the fluctuation spectrum in the $q \sim 1$ region has a high degree of degeneracy, allowing it to respond differently to different kinds of internal and external forces and sources. For instance, this facilitates quasi-interchange dynamics and associated flux pumping \cite{Wesson86, Krebs17}, and supports nearby quasi-equilibria with a tilted axis or helical substructures \cite{Wesson86, Weller87, Cooper10, Cooper11, Wingen18, Yun12, Bierwage15a, Adulsiriswad25}. When $q \sim 1$, the effect of small differences between the magnetic and kinetic helicities are also enhanced, facilitating energy- and pitch-selective confinement \cite{Bierwage22b}.

It has also been recognized very early that instabilities in regions of weak magnetic shear can have long-range effects in toroidal geometric \cite{Zakarov78}. Pressure-gradient-driven instabilities centered in the weak-shear region near any low-rational value of $q$ (such as $q \sim 2$) are often referred to collectively as ``infernal modes'' \cite{Manickam87}, which include also resistive modes involving magnetic reconnection \cite{Charlton89} and continue to be an active area of research and modeling \cite{CosteSarguet24}.

Here, we report the finding of another peculiar phenomenon in a numerically simulated tokamak with centrally flat $q \sim 1$: slow magnetoacoustic pulsations (SMAP) in the form of an oscillatory quasi-mode on the central plateau of the acoustic continuum, $\omega_{\rm S} = k_\parallel c_{\rm s} \approx {\rm const}$., with parallel wavenumber $k_\parallel$, sound speed $c_{\rm S}$, and dominant poloidal and toroidal mode numbers $m/n=1/1$. The core-localized pulsations continue after their quasi-linear saturation and can couple to distant $q = 2,3,...$ rational surfaces, where they drive magnetic reconnection and give rise to pulsating macroscopic magnetic islands.

Our parameter scans show that SMAPs are facilitated by low thermal conductivity, which allows thermal misbalance to persist. This is reminiscent of the back-reaction of wave-induced thermal misbalance on MHD waves (slow magnetoacoustic and entropy waves) and resulting pulsations in the solar corona as described, for instance, by Kolotkov {\it et al}.\ \cite{Kolotkov23}, who recently examined the effect of parallel heat conductivity and pointed out similarities between the driven coronal plasma and a laser's active gain medium. Potentially related phenomena seem to exist also in Earth's magnetosphere, where intricacies pertaining to Alfv\'{e}n-acoustic couplings due to the plasma's geometry and inhomogeneity, and the resulting ballooning-type nature of modes were discussed by Rubtsov {\it et al}.\ \cite{Rubtsov18}.

Similar to ballooning and other interchange modes, the SMAPs in our simulations can be energized by gradients in the ambient pressure. Our setup is ideally stable (cf.~Fig.~A.1 of Ref.~\cite{Bierwage26a}), but SMAPs can be destabilized when ion-viscous damping is sufficiently weak and electric resistivity $\eta$ is sufficiently high. With increasing growth rate $\gamma$, the oscillation frequency $\omega = \omega_{\rm pulse}/2$ decreases as $\omega \approx \omega_{\rm S} - \gamma$, and growth-rate's resistivity dependence seems to approach the scaling $\gamma \propto \eta^{5/3}$. The latter is an indication of magnetic reconnection. Although no reconnection could be detected in the central $q \sim 1$ region, it is possible that the minute changes in the magnetic topology are undone by the loop voltage, whose parallel electric field $E_{\parallel{\rm src}} = -\eta J_{\rm ref}$ acts to restore the ambient current density $J_{\rm ref} \propto 1/q$ by counter-acting its dissipation.

However, when the simulation domain encompasses rational surfaces with $q = 2,3,4,5$, magnetic reconnection is observed at those distant radii all the way to the plasma edge. The simulations where this is seen have a resistive layer width of $\delta r_\eta \approx 0.03 R_0 \approx 0.1a$ on the time scale of the slow magnetoacoustic oscillations with frequency $\omega_{\rm S}/\omega_{\rm A0} \sim 0.008$, which allows the loop voltage to undo the reconnection during the course of each pulse, so that the islands also pulsate, disappearing and reappearing periodically, with their X- and O-point locations interchanged alternatingly.

From the theoretical point of view, the fact that SMAPs oscillate with period $2\pi/\omega_{\rm S}$ even when growing or decaying exponentially is remarkable because --- in the absence of differential rotation and resonant wave-particle interactions --- a conventional linear ideal MHD eigenmode is expected to be either purely growing or purely oscillating \cite{Frieman60}. We are not aware of reported cases where the addition of visco-resistive or thermal diffusion violated this theorem. The reasons by which SMAPs become an exception to this fundamental theorem will be discussed after reporting our results.

The simulation model is described in Section~\ref{sec:model}. Results of parameter scans are presented in Section~\ref{sec:results}, followed by a summary and discussion in Section~\ref{sec:discussion}, and conclusions and an outlook in Section~\ref{sec:conclusion} with motivation for further study using kinetic models. One such avenue for future work with kinetic models are preliminary observations of an Alfv\'{e}nic counterpart of SMAPs, which we call ``low-frequency compressible Alfv\'{e}nic pulsations (LCAP)''. LCAPs satisfy $\omega \approx \omega_{\rm A} = k_\parallel v_{\rm A}$ with Alfv\'{e}n speed $v_{\rm A}$, are less dependent on thermal misbalance, and are marginally stable for $\eta=0$. Resonant drive may be required for LCAPs to grow, and collisionless mechanisms may be needed to facilitate forced reconnection \cite{Hahm85} similar to that seen for SMAPs.

\section{Model}
\label{sec:model}

We use the visco-resistive single-fluid full MHD module of the MHD-PIC hybrid code {\tt MEGA} \cite{Todo98, Todo05, Todo25}, which evolves the mass flux vector ${\bm \Gamma}_{\rm M}(t) = \rho_{\rm M}{\bm u}$, scalar mass density $\rho_{\rm M}(t)$, scalar thermal pressure $P(t)$, and magnetic vector ${\bm B}(t)$ by solving the momentum equation, continuity equation, thermodynamic equation of state, and magnetic induction equation:
\begin{subequations}
	\begin{align}
		\partial_t{\bm \Gamma}_{\rm M} =& -\nablab\cdot({\bm \Gamma}_{\rm M} {\bm u}) - \nablab P + {\bm J}\times{\bm B} \nonumber
		\\
		&- \nablab\times(\eta_{\rm is} \rho_{\rm M}\nablab\times{\bm u}) + \tfrac{4}{3} \nablab(\eta_{\rm ib} \rho_{\rm M} \nablab\cdot{\bm u}),
		\label{eq:model_cc_mom}
		\\
		\partial_t\rho_{\rm M} =& -\nablab\cdot{\bm \Gamma}_{\rm M} + \chi_{\rm th}\nabla^2 (\rho_{\rm M}-\rho_{\rm M,ref}),
		\label{eq:model_cc_den}
		\\
		\partial_t P =& -\nablab\cdot(P{\bm u}) - (\Gamma - 1)P\nablab\cdot{\bm u} + \chi_{\rm th}\nabla^2(P - P_{\rm ref}), \nonumber
		\\
		& + (\Gamma - 1)\left[\eta_{\rm is}\rho_{\rm M}(\nablab\times{\bm u})^2 + \tfrac{4}{3}\eta_{\rm ib}\rho_{\rm M}(\nablab\cdot{\bm u})^2 \right. \nonumber \\
		& \qquad \qquad\;\; \left. + \eta_{\rm e} {\bm J}\cdot({\bm J} - {\bm J}_{\rm ref})\right],
		\label{eq:model_cc_pre}
		\\
		\partial_t{\bm B} =& -\nablab\times{\bm E}.
		\label{eq:model_cc_bfield}
	\end{align}\vspace{-0.5cm}
	\label{eq:model_cc}
\end{subequations}

\noindent {\tt MEGA} works with right-handed cylinder coordinates $(R,\varphi,z)$ and uses 4th-order finite-differencing in space. The time derivative $\partial_t \equiv \partial/\partial t$ is inverted by 4th-order Runge-Kutta integration. Each Runge-Kutta integration step is followed by the computation of the electric current density ${\bm J}$, electric field vector ${\bm E}$, and fluid velocity vector ${\bm u}$ from the algebraic equations
\begin{subequations}
	\begin{align}
		\mu_0 {\bm J} =&\, \nablab\times{\bm B},
		\label{eq:model_cc_cur}
		\\
		{\bm E} =&\, -{\bm u}\times{\bm B} + \eta_{\rm e}({\bm J} - {\bm J}_{\rm ref}),
		\label{eq:model_cc_efield}
		\\
		{\bm u} \equiv&\, {\bm \Gamma}_{\rm M}/\rho_{\rm M};
		\label{eq:model_cc_u}
	\end{align}
	\label{eq:model_cc_alg}
\end{subequations}

\noindent where (\ref{eq:model_cc_cur}) is Amp\`{e}re's law with vacuum magnetic susceptibility $\mu_0$, and (\ref{eq:model_cc_efield}) is known as Ohm's law.

The reference state shown in Fig.~\ref{fig:01_setup} is taken to be a toroidally axisymmetric ($\nablab\varphi\cdot\nablab = 0$) stationary (${\bm u}_{\rm ref} = 0$) model of a KSTAR tokamak plasma \cite{Lee23, Ko24} consisting of deuterium (with particle mass $M_{\rm D}$) that has nonzero number density and pressure everywhere ($N_{\rm ref} \equiv \rho_{\rm M,ref}/M_{\rm D} \neq 0$, $P_{\rm ref} \neq 0$). Given as inputs the shape of the plasma boundary shown in Fig.~\ref{fig:01_setup}(a), one of the pressure profiles $P_{\rm ref}$ or $P_{\rm flat}$ in panel (b), and the safety factor profile $q$ in panel (c), the ideal MHD equilibrium code {\tt CHEASE} \protect\cite{Luetjens96} was used to compute the poloidal flux $2\pi\Psi$ and covariant toroidal component $I(\Psi) = \partial_\varphi{\bm r}\cdot{\bm B}_{\rm ref} = R B_{\rm tor,ref}$ of the ambient magnetic field ${\bm B}_{\rm ref} = \nablab\Psi\times\nablab\varphi + I\nablab\varphi$. The magnetic axis at major radius $R_0$ and height $z_0$ is surrounded by nested toroidal magnetic surfaces ($\Psi = {\rm const}$.). The rational surfaces where the magnetic field lines close on themselves after one poloidal turn and $q = 2,3,4,5$ toroidal turns are plotted as white dotted lines in Fig.~\ref{fig:01_setup}(a). The normalized poloidal flux $\psi \equiv (\Psi - \Psi_0)/(\Psi_a - \Psi_0)\in [0,1]$ relative to its on-axis and boundary values, $\Psi_0$ and $\Psi_a$, is used as a minor radial coordinate; either directly or in terms of its square root, $\psi^{1/2}$, which resembles more closely the minor radius in real space.

Physics that break ideal MHD constraints are modeled by diffusion terms with scalar coefficients measuring thermal diffusion ($\chi_{\rm th}$), electric resistivity ($\eta_{\rm e}$), and the ion fluid's shear and bulk viscosities ($\eta_{\rm is}$, $\eta_{\rm ib}$). Here, we let $\eta_{\rm i} = \eta_{\rm is} = \eta_{\rm ib}$ and assume that all diffusion coefficients are spatially uniform and constant in time. The dissipation of the reference state is counteracted by source terms, such as $-\eta_{\rm e}{\bm J}_{\rm ref}$ in Ohm's law (\ref{eq:model_cc_efield}), which represents the loop voltage in an actual tokamak plasma. The above model can be shown to formally conserve energy, at least in the absence of the density diffusion term $\chi_{\rm th}\nablab^2 (\rho_{\rm M}-\rho_{\rm M,ref})$ in Eq.~(\ref{eq:model_cc_den}). While helping to maintain numerical stability at small scales, our parameter scans indicate that this term seems to have a negligible effect on the long-wavelength modes that we study here.

We normalize time by the on-axis Alfv\'{e}n time $\tau_{\rm A0} \equiv R_0/v_{\rm A0}$, velocities by the on-axis Alfv\'{e}n speed $v_{\rm A0} \equiv B_0/\sqrt{\mu_0\rho_{\rm M,0}}$, the mass density and magnetic field by their on-axis values $\rho_{\rm M,0}\equiv\rho_{\rm M,ref}(R_0,z_0)$ and $B_0 \equiv B_{\rm ref}(R_0,z_0)$, and spatial gradients by $R_0$. The normalized diffusion coefficients are then
\begin{equation}
	\eta \equiv \frac{\eta_{\rm e}}{\mu_0 v_{\rm A0} R_0}, \quad \nu \equiv \frac{\eta_{\rm i}}{v_{\rm A0} R_0}, \quad \chi \equiv \frac{\chi_{\rm th}}{v_{\rm A0} R_0}.
\end{equation}

\noindent The toroidal versions (on the $R_0$ scale) of the Lundquist and Reynolds number are then $S \equiv \eta^{-1}$ and ${\rm Re} \equiv \nu^{-1}$.

Before simulating the entire plasma in Fig.~\ref{fig:01_setup}, we will use smaller simulation domains covering only 50\% or 75\% of the poloidal flux space. The corresponding plasma boundaries, where we apply a non-slip boundary condition (${\bm u} = 0$), are indicated by dashed red lines in Fig.~\ref{fig:01_setup}(a)--(c). The $50\%$ domain does not have any integer rational surfaces, the $75\%$ domain has only $q=2$, and the full domain has integer resonances $q = 2,3,4,5$. The number of grid points used in the respective simulations were $N_R\times N_z\times N_\varphi = 256\times 256\times 16$, $512\times 512\times 16$ and $1040\times 1040\times 16$. Tests with higher resolutions were also performed to ascertain the reliability of the results.

Before beginning the actual simulation, {\tt MEGA} advances Eqs.~(\ref{eq:model_cc}) by one time step and saves whatever prompt relaxation $\delta \tilde{\bm F} \equiv \{\delta\tilde{\bm\Gamma}_{\rm M},\delta\tilde{\rho}_{\rm M},\delta\tilde{P},\delta\tilde{\bm B}\}$ occurs due to inconsistencies between the discretized ideal MHD equilibrium of {\tt CHEASE} and the unknown exact equilibrium of the the visco-resistive full MHD equations (\ref{eq:model_cc}) on the actual computational mesh. When the simulation begins at time $t=0$, {\tt MEGA} introduces an arbitrary initial perturbation $\delta{\bm F}(t=0)$ with user-defined amplitude and then proceeds to simulate the plasma response by solving Eqs.~(\ref{eq:model_cc}) at suitable time steps $\Delta t$ satisfying the Courant-Friedrichs-Lewy (CFL) condition, while subtracting the above prompt relaxation fields like
\begin{equation}
	\delta {\bm F}(t+\Delta t) = \delta{\bm F}(t) + \Delta t\times\left\{\text{[r.h.s.\ of Eq.~(\ref{eq:model_cc})]} - \delta \tilde{\bm F}\right\}.
\end{equation}

During the simulation, we apply a Fourier filter with basis ${\rm exp}(-in\varphi)$ in the toroidal direction and retain only the components with toroidal mode numbers $|n| = 0,1$. For data analysis and visualization, the fields are decomposed in the Fourier basis $\exp(im\vartheta - in\varphi - i\omega t)$ using right-handed straight-field-line toroidal flux coordinates $(\psi,\varphi,\vartheta)$ with poloidal angle $\vartheta$. Using the Fourier components of the normalized radial MHD velocity $\delta\hat{u}_{\rm rad} \equiv \hat{\bm e}_r\cdot\delta{\bm u}/v_{\rm A0}$, the fluctuating electrostatic potential $\delta\hat{\Phi} = \delta\Phi/(a v_{\rm A0} B_0)$ is estimated as
\begin{equation}
	\delta\hat{\Phi}_n(r|m\neq 0) \approx -i\frac{\hat{r}}{m}[\delta\hat{u}_{\rm rad}]_n(r|m\neq 0),
	\label{eq:epot}
\end{equation}

\noindent where $\hat{r}(\psi) = r/a \approx \psi_{\rm tor}^{1/2}$ is the volume-averaged minor radius, which is similar to the square root of the normalized toroidal magnetic flux. The MHD fluctuation energy of toroidal Fourier component $n$ is measured as
\begin{align}
	W_n \equiv \sum\limits_{n = \pm|n|} \left. \int{\rm d}^3x\, \right(&\frac{\mu_0}{2}\left|\rho_{\rm M}^{1/2}\delta{\bm u}\right|_n^2 + \frac{\mu_0\delta P_n}{\Gamma - 1} \nonumber \\
	&\left. + \frac{1}{2}\delta{\bm B}_n\cdot(\delta{\bm B}_n + 2{\bm B}_{\rm ref})\right),
	\label{eq:wn}
\end{align}\vspace{-0.4cm}

\noindent where the last term is identical to $\tfrac{1}{2}(B^2 - B_{\rm ref}^2)$.

\begin{figure*}
	[tb!]\vspace{-2.25cm}
	\centering
	\includegraphics[width=0.96\textwidth]{\figures/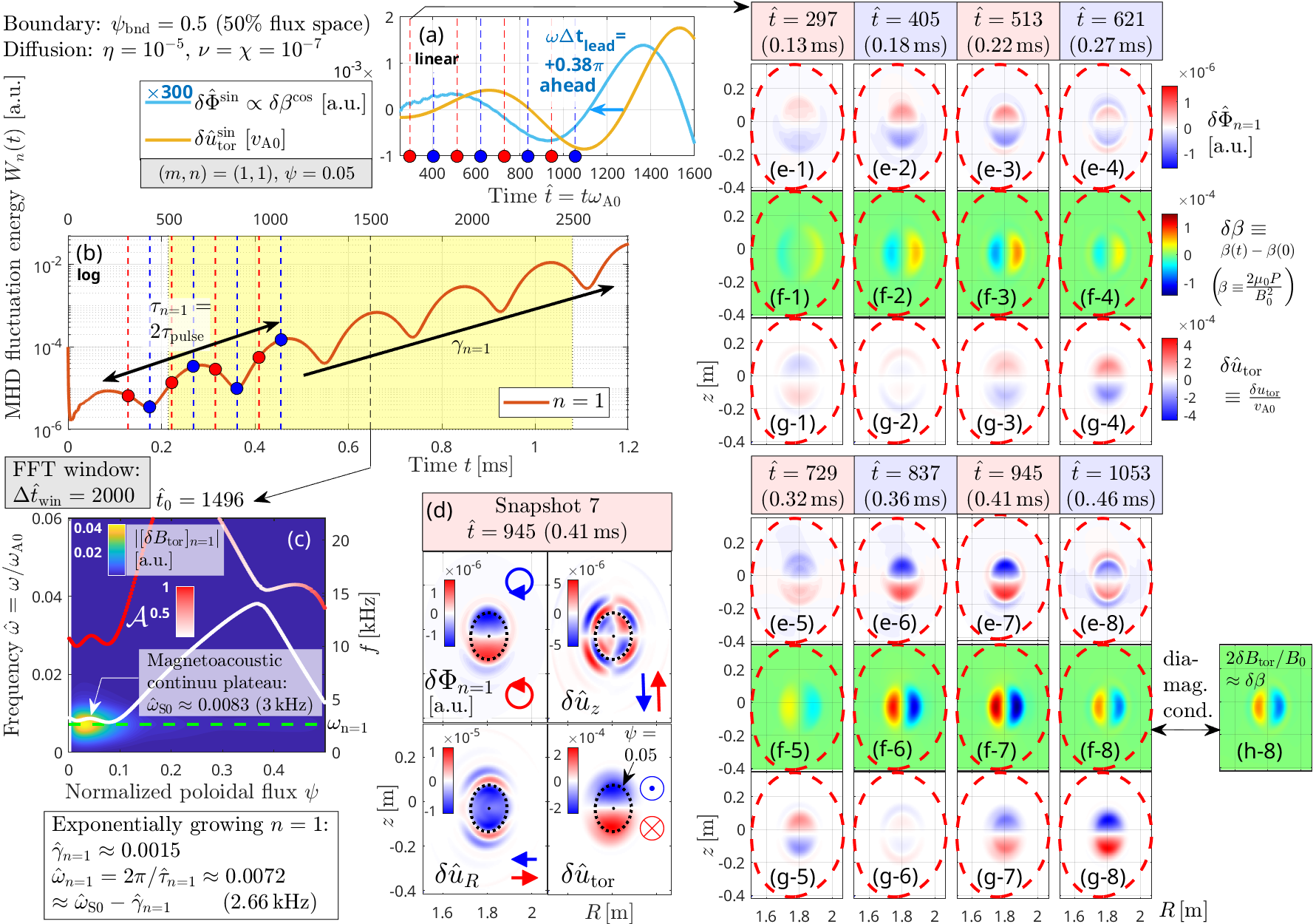}\vspace{-0.1cm}
	\caption{Exponential growth of slow magnetoacoustic pulsations (SMAPs) in a simulation with high resistivity ($\eta=10^{-5}$), and low viscosity and diffusivity ($\nu = \chi = 10^{-7}$). The specific heat ratio is $\Gamma = 5/3$. Here, the core-localized quasi-mode is isolated by using a reduced simulation domain, covering only the inner $50\%$ of flux space ($\psi_{\rm bnd} = 0.5$), just inside the $q=2$ surface, so that there are no integer resonances. Panel (a) shows the time traces of the $(m,n)=(1,1)$ Fourier components of the electrostatic potential $\delta\hat{\Phi}^{\rm sin}$ and toroidal MHD velocity $\delta\hat{u}_{\rm tor}^{\rm sin}$ at $\psi = 0.05$ near the plasma center. $\delta\hat{\Phi}^{\rm sin}$ was scaled by a factor $300$ to be visible on the same scale as $\delta\hat{u}_{\rm tor}^{\rm cos}$. In this particular case, the sine component of the potential and the cosine component of the normalized pressure perturbation happen to have nearly identical phases, so only one curve is plotted for $\delta\hat{\Phi}^{\rm sin}(t) \propto \delta\beta^{\rm cos}(t)$, and both are leading ahead of $\delta\hat{u}_{\rm tor}^{\rm sin}(t)$ with a phase difference $\omega\Delta t_{\rm lead} \approx +0.38\pi$. Panel (b) shows on a logarithmic scale the time trace of the magnetic fluctuation energy $W_{n=1}(t)$ as defined in Eq.~(\protect\ref{eq:wn}). The upper axis shows time $\hat{t} \equiv t\omega_{\rm A0}$ in units of Alfv\'{e}n time, and the lower axis shows time in milliseconds. The evolution after saturation is shown in Fig.~\protect\ref{fig:03_flux0.50cut2_577-evol}. For the Fourier time window $\hat{t}_0 \pm \Delta\hat{t}_{\rm win}/2$ indicated by the yellow shaded area in (b), panel (c) shows the spectrogram of the toroidal magnetic fluctuations $|[\delta B_{\rm tor}]_{n=1}|(\omega,\psi)$ (parula-colored), overlaid with the MHD continua from Fig.~\protect\ref{fig:01_setup}(d). The horizontal green dashed line in panel (c) indicates the measured frequency $\omega_{n=1} = 0.0072\,\omega_{\rm A0} = 2\pi\times 2.66\,{\rm kHz}$ of the oscillations seen in panel (a). Note that each oscillation period $\tau_{n=1}$ consists of two SMAP periods $\tau_{\rm pulse} = \tau_{n=1}/2$ as indicated by the double arrow in panel (b). As noted in the box below panel (c), $\omega_{n=1}$ is down-shifted relative to the slow magnetoacoustic frequency plateau $\omega_{\rm S0}$ by an amount similar to the exponential growth rate $\gamma_{n=1}$. On the right-hand side of the figure, panel groups (e)--(g) each show a series of 8 snapshots of the electrostatic potential $\delta\Phi_{n=1}(R,z)$, the normalized pressure fluctuation $\delta\beta(R,z)$, and the toroidal MHD velocity field $\delta u_{\rm tor}$. The red and blue circles in panels (a) and (b) indicate the snapshot times. For snapshot 7, panel set (d) shows the orientations of the ${\bm E}\times{\bm B}$ vortex flows represented by the contours of $\delta\Phi_{n=1}(R,z)$ as well as the orientations of radial, vertical and toroidal MHD flows. Note that the oscillation magnitude of $|\delta u_{\rm tor}|$ is $20$ to $40$ times larger than $|\delta u_{\rm pol}| \equiv |\delta u_R^2 + \delta u_z^2|^{1/2}$. For the last snapshot of $\delta\beta$ in panel (f-8), an additional panel (h-8) shows that $2\delta B_{\rm tor}/B_0$ is effectively identical to $\delta\beta$, so that the diamagnetic condition (\protect\ref{eq:diamag_condition}) for slow magnetoacoustic waves is satisfied to a high level of accuracy.}
	\label{fig:02_flux0.50cut2_577-evol-exp}\vspace{-0.2cm}
\end{figure*}

\section{Results}
\label{sec:results}

For a configuration very similar to that in Fig.~\ref{fig:01_setup}, the existence of a core-localized coherent quasi-mode with Alfv\'{e}nic polarization was demonstrated numerically in Ref.~\cite{Bierwage26a} using an electromagnetic antenna model. In those simulations, the core-localized quasi-mode formed near or below the Alfv\'{e}nic continuum branch with dominant poloidal and toroidal mode numbers $m/n=1/1$, whose central plateau can be seen as a nearly horizontal red line in Fig.~\ref{fig:01_setup}(d) around $\rho \lesssim 0.3$. Since $q(\rho \lesssim 0.3)$ is very close to (slightly above) unity, the $m/n = 1/1$ continua of shear Alfv\'{e}n and acoustic branches below the Alfv\'{e}n-acoustic gap have a small but nonzero frequency in the central core. For the Alfv\'{e}nic branch, appearing red in Fig.~\ref{fig:01_setup}(d), the central plateau frequency is about $10\,{\rm kHz}$. For the acoustic branch, appearing white in Fig.~\ref{fig:01_setup}(d), the central plateau frequency is $3\,{\rm kHz}$ in the present simulation setup with on-axis toroidal beta $\beta_0 = 2\mu_0 P_{\rm ref,0}/B_0^2 \approx 4\%$ and specific heat ratio $\Gamma = 5/3$.

In Fig.~C.1 of Ref.~\cite{Bierwage26a} it was shown that the antenna-driven quasi-mode exhibited long-lasting electromagnetic pulsations before settling in a quasi-steady state when resistive and viscous dissipation as well as thermal diffusivity were weak: $\eta = \nu = \chi = 10^{-7}$. The relaxation time scale was significantly reduced when raising some or all of these values to $10^{-6}$, while the saturation amplitude and mode structure remained about the same. Finally, when taking $\eta = \nu = \chi = 10^{-5}$, the saturation amplitude was noticeably reduced and the mode structure broadened by the enhanced diffusion.

Here, we report results of additional curiosity-driven parameter scans, which revealed that high resistivity ($\eta > 10^{-6}$) in combination with low viscosity and thermal diffusivity ($\nu,\chi < 10^{-6}$) destabilizes a quasi-mode on the $m/n=1/1$ acoustic continuum plateau (at $3\,{\rm kHz}$ around $\rho \lesssim 0.3$ in Fig.~\ref{fig:01_setup}(d)). Even without antenna drive and independently of the initial perturbation amplitude, this quasi-mode grows exponentially on average while pulsating periodically. We will later show that these core-localized slow magnetoacoustic pulsations (SMAP) can couple to distant rational surfaces $q=2,3,...$, which then undergo magnetic reconnection. Before that, we study the properties of SMAPs in the absence of rational surfaces by truncating the simulation domain at $50\%$ of the poloidal flux space, placing the non-slip boundary at the innermost red dashed line in Fig.~\ref{fig:01_setup}(a)--(c), just inside the $q=2$ surface.

\subsection{Core-localized SMAP in isolation}
\label{sec:results_0.5}

\begin{figure*}
	[tb!]\vspace{-0.7cm}
	\centering
	\includegraphics[width=0.96\textwidth]{\figures/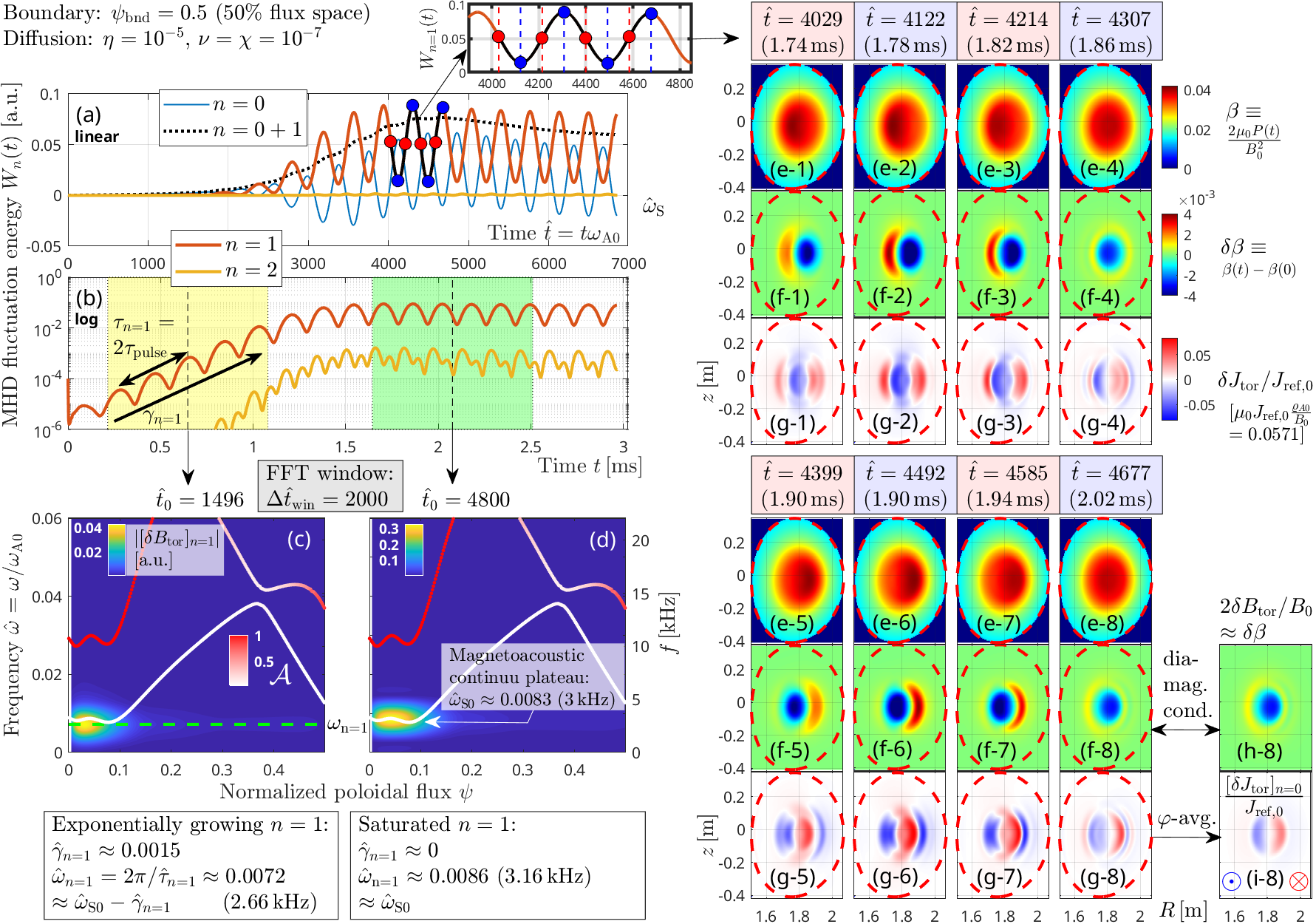}
	\caption{Growth, saturation and long-time evolution of slow magnetoacoustic pulsations (SMAPs). Continuation of Fig.~\protect\ref{fig:02_flux0.50cut2_577-evol-exp}. Panels (a) and (b) show --- respectively on a linear and logarithmic scale --- the time traces of the magnetic fluctuation energies $W_n(t)$ as defined in Eq.~(\protect\ref{eq:wn}) for $n=0$ (blue), $n=1$ (brown), and $n=2$ (orange, negligibly small). The black dotted curve is $W_{n=0} + W_{n=1}$. For the Fourier time windows $\hat{t}_0 \pm \Delta\hat{t}_{\rm win}/2$ indicated by yellow- and green-shaded areas in (b), panels (c) and (d) show spectrograms of the toroidal magnetic fluctuations $|[\delta B_{\rm tor}]_{n=1}|(\omega,\psi)$ (parula-colored), overlaid with the MHD continua from Fig.~\ref{fig:01_setup}(d). Note that, after saturation ($\gamma \rightarrow 0$), the oscillation frequency $\omega_{n=1}$ matches the magnetosonic continuum plateau frequency $\omega_{\rm S0}$. On the right-hand side, panel groups (e)--(g) each show a series of 8 snapshots of the normalized pressure field $\beta(R,z)$, its fluctuating component $\delta\beta$, and the toroidal current density fluctuations $\delta J_{\rm tor}$. The red and blue circles in panel (a) indicate the snapshot times. For the last snapshot of $\delta\beta$ in panel (f-8), an additional panel (h-8) shows that $2\delta B_{\rm tor}/B_0$ is effectively identical to $\delta\beta$, so that the diamagnetic condition (\protect\ref{eq:diamag_condition}) for slow magnetoacoustic waves is satisfied to a high level of accuracy. For the last snapshot of $\delta J_{\rm tor}$ in panel (g-8), an additional panel (i-8) shows the axisymmetric ($\varphi$-averaged) component $[\delta J_{\rm tor}]_{n=0}$, which only oscillates in magnitude, not in direction. Additional time traces of $\delta{\bm u}$ and $\delta B_{\rm tor}$ measured at the location of the slow magnetoacoustic quasi-mode ($\psi \approx 0.05$) are shown in Fig.~\protect\ref{fig:04_flux0.50_sat-trace}.}
	\label{fig:03_flux0.50cut2_577-evol}
\end{figure*}

Figs.~\ref{fig:02_flux0.50cut2_577-evol-exp} and \ref{fig:03_flux0.50cut2_577-evol} summarize results of a simulation covering the inner 50\% of poloidal flux space (just inside the $q=2$ surface). We use a relatively high resistivity $\eta = 10^{-5}$ and low values of the viscosity and thermal diffusivity $\nu = \chi = 10^{-7}$, which will be our default settings in this paper. Times and frequencies will be given in units of Alfv\'{e}n time $\tau_{\rm A0} \equiv \omega_{\rm A0}^{-1} \equiv R_0/v_{\rm A0}$ as $\hat{t} \equiv t\omega_{\rm A0}$ and $\hat{\omega}_{\rm A0} = \omega/\omega_{\rm A0}$, often together with the corresponding values in milliseconds (ms) and kilohertz (kHz).

Starting from an initial perturbation amplitude corresponding to $W_{n=1}(t=0) = 10^{-4}$, the MHD fluctuation energy $W_{n=1}(t)$ in Figs.~\ref{fig:02_flux0.50cut2_577-evol-exp}(b) and \ref{fig:03_flux0.50cut2_577-evol}(b) can be seen to drop abruptly to about $10^{-6}$ before it grows in a pulsating manner but exponentially on average during the first $3000$ Alfv\'{e}n times ($1.3\,{\rm ms}$). The properties of this pulsating mode during the phase of net exponential growth are summarized in Fig.~\ref{fig:02_flux0.50cut2_577-evol-exp}.

It must be emphasized that one full oscillation period $\tau_{n=1}$ of the wave's $n=1$ Fourier component in Fig.~\ref{fig:02_flux0.50cut2_577-evol-exp}(a) comprises two pulsations $\tau_{\rm pulse}$ of $W_{n=1}$ in Fig.~\ref{fig:02_flux0.50cut2_577-evol-exp}(b) (the same holds after saturation in Fig.~\ref{fig:03_flux0.50cut2_577-evol}). The frequency can be determined by measuring the pulsation period in panel (b) and calculating $\omega_{n=1} = 2\pi/\tau_{n=1} = \pi/\tau_{\rm pulse}$, or from the peak of the FFT spectrogram in panel (c). Both methods give consistent results.

Fig.~\ref{fig:02_flux0.50cut2_577-evol-exp}(c) shows that the exponentially growing instability is located around $0 \lesssim \psi \lesssim 0.1$ near the plasma center and oscillates at a frequency $\hat{\omega}_{n=1} \approx 0.0072$ ($2.86\,{\rm kHz}$), slightly below the central plateau frequency $\hat{\omega}_{\rm S0} \approx 0.0083$ ($3\,{\rm kHz}$) of the slow magnetoacoustic continuum's $(m,n)=(1,1)$ branch.

\begin{figure*}
	[tb!]\vspace{-2.0cm}
	\centering
	\includegraphics[width=0.48\textwidth]{\figures/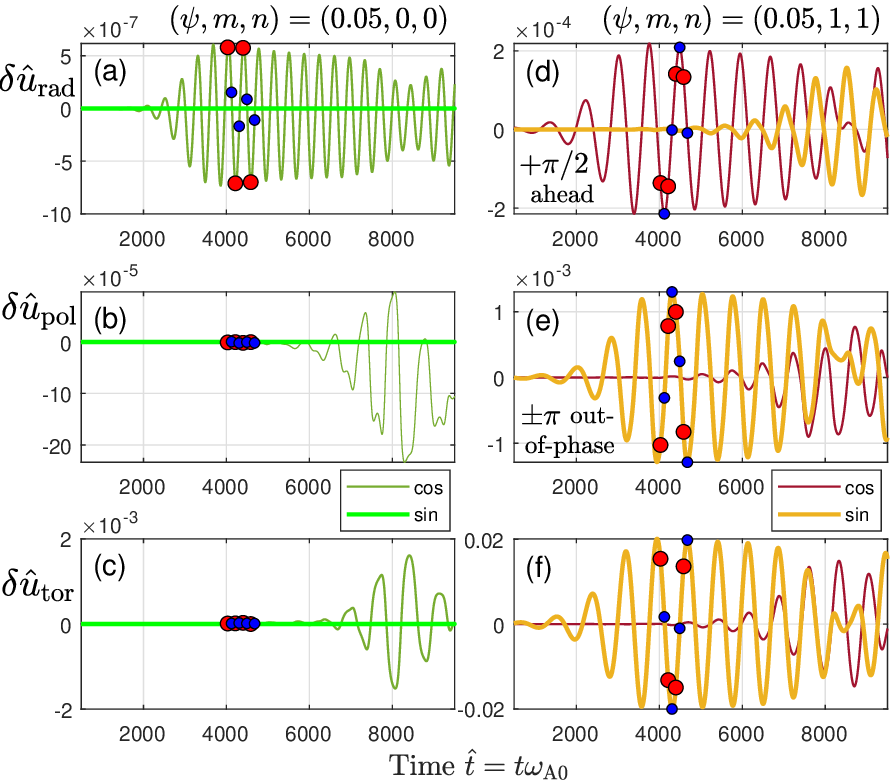}
	\includegraphics[width=0.48\textwidth]{\figures/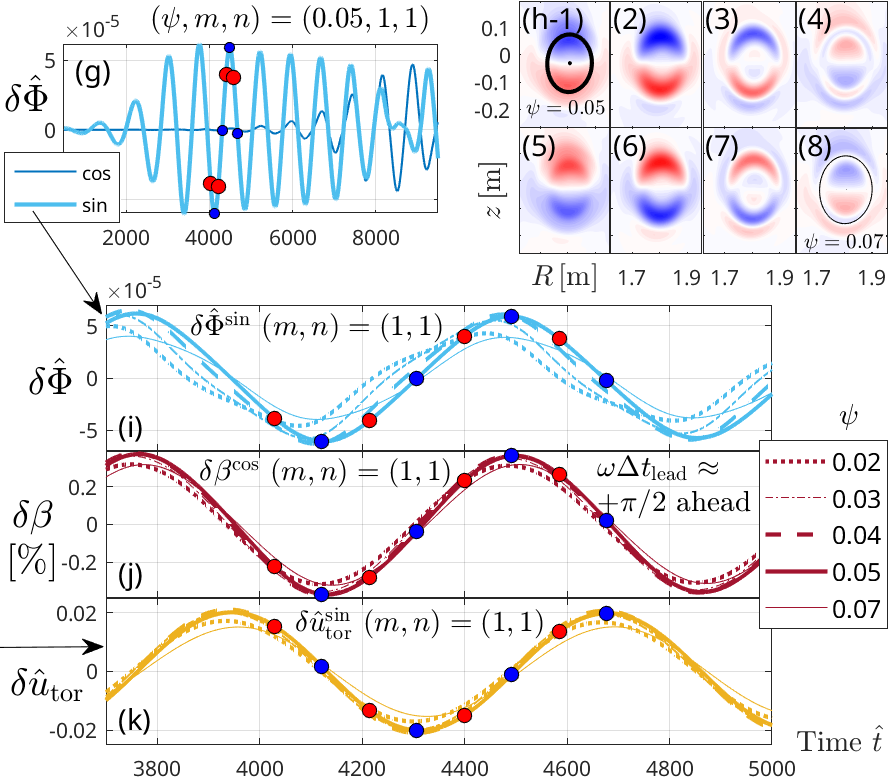}
	\caption{Local waveforms in the saturated regime of Fig.~\protect\ref{fig:03_flux0.50cut2_577-evol}. The left-hand side of this figure shows the time traces of the $(m,n) = (0,0)$ and $(1,1)$ Fourier components of the MHD velocity components $\delta u_{\rm rad}$ (a,d), $\delta u_{\rm pol}$ (b,e), and $\delta u_{\rm tor}$ (c,f) at radius $\psi = 0.05$, where the core-localized slow magnetoacoustic $n=1$ quasi-mode is located. The $(m,n)=(0,1)$ and $(1,-1)$ components are 1 or 2 orders of magnitude smaller than $(1,1)$ and therefore not shown here. Dark colors represent the cosine and light colors the sine component. On the right-hand side, panel (g) shows $\delta\Phi^{\rm sin}_{1,1} \propto [\delta u_{\rm rad}]_{1,1}^{\rm cos}$, and panel set (h) shows 8 snapshots of $\delta\Phi_{n=1}(R,z)$ taken at the times indicated by red and blue circles (same as in Fig.~\protect\ref{fig:03_flux0.50cut2_577-evol}). Panels (i)--(k) show enlargements of the local signals $\delta\Phi^{\rm sin}(t)$, $\delta\beta^{\rm cos}(t)$, $\delta u_{\rm tor}^{\rm sin}(t)$ around $\hat{t} \approx 4500$ ($2\,{\rm ms}$) measured at 5 radii in the range $0.02 \leq \psi \leq 0.07$.}
	\label{fig:04_flux0.50_sat-trace}
\end{figure*}

The snapshots on the right-hand side of Fig.~\ref{fig:02_flux0.50cut2_577-evol-exp} show the evolution of the spatial structure of $\delta\Phi_{n=1}$, $\delta\beta$ and $\delta u_{\rm tor}$ in the poloidal plane $(R,z)$ at $\varphi = 0$. The snapshot times are indicated by red and blue circles in panels (a) and (b). Panel (d) clarifies the relation between the red and blue colors of the ${\bm E}\times{\bm B}$ vortices in $\delta\Phi_{n=1}(R,z)$ as well as orientation of the toroidal flow $\delta u_{\rm tor}$. Three noteworthy observations can be made in Fig.~\protect\ref{fig:02_flux0.50cut2_577-evol-exp} for the phase of exponential growth:
\begin{enumerate}
	\item[(i)]  The time traces in panel (a) and the snapshots in panel sets (e) and (g) show that the oscillations of $\delta\Phi$ precede those of $\delta u_{\rm tor}$ by $\omega\Delta t_{\rm lead} \approx +0.38\pi$. Moreover, $\delta\Phi^{\rm sin}$ and $\delta \beta^{\rm cos}$ happen to have identical phases. However, the relative phasing between these three quantities depends on the growth rate $\gamma$ (or on the parameters that control it, such as $\eta$).
	\item[(ii)]  The snapshots of $\delta\Phi_{n=1}(R,z)$ and $\delta u_{\rm tor}(R,z)$ in Fig.~\ref{fig:02_flux0.50cut2_577-evol-exp}(e,g) show that the mode's phase reversal between subsequent pulses does not proceed in a uniform manner (as would be expected for an eigenmode), but involves radial corrugation and pattern propagation. These transients are more prominent in the weaker Alfv\'{e}nic (electromagnetic) component represented by $\delta\Phi_{n=1}$ than in the stronger acoustic component represented by $\delta u_{\rm tor}$.
	\item[(iii)]  Panels (f-8) and (h-8) show that the normalized perturbations of the thermal pressure and toroidal magnetic field are very similar: $\delta\beta \approx 2\delta \hat{B}_{\rm tor} \equiv 2\delta B_{\rm tor}/B_0$. This is representative for all snapshots (also later, in the saturated regime).
\end{enumerate}

\noindent The phasing between the three fields mentioned in item (i) seems to be related to the quasi-mode's ability to grow exponentially by extracting energy from the thermal pressure gradient in the $q \sim 1$ region. This will be discussed in more detail later. Observation (ii) confirms that we are observing a quasi-mode rather than an eigenmode. Observation (iii) means that the pressure and magnetic perturbations satisfy what in the space plasma literature is known as the diamagnetic condition\footnote{The diamagnetic condition (\protect\ref{eq:diamag_condition}) is usually invoked for short wavelengths \protect\cite{MaHirose14,Rubtsov18}, but it is satisfied also by our long-wavelength modes.}
\begin{equation}
	\mu_0 \delta P + B\delta B_\parallel = 0.
	\label{eq:diamag_condition}
\end{equation}

\noindent Near the magnetic axis of our KSTAR tokamak with $B_{\rm tor} \equiv \hat{\bm e}_\varphi\cdot{\bm B} \approx -B_0$, this translates to $\delta\beta = - 2\delta\hat{B}_\parallel \approx 2\delta\hat{B}_{\rm tor}$. This means that we are dealing with a slow mode, while the fast waves maintain a state of near-MHD-force-balance at all times almost instantaneously during the slow oscillations that are seen here.

Let us now proceed to the saturated regime, results for which are summarized in Fig.~\ref{fig:03_flux0.50cut2_577-evol}. SMAPs in this case saturate at about $\hat{t} \approx 3500$ ($1.5\,{\rm ms}$), and Fig.~\ref{fig:03_flux0.50cut2_577-evol}(b) shows how the pulsations continue with more or less constant amplitude. Most notably, we observe the following:
\begin{enumerate}
	\item[(iv)] The $n=0$ and $n=1$ components always pulsate in unison and have comparable amplitudes, but are $180$ degrees out-of-phase.
	\item[(v)] A close inspection of the pulsation cycle $\tau_{\rm pulse}$ in panel (b) reveals a slight speed-up compared to the phase of exponential growth. This is confirmed by the spectrograms in panels (c) and (d), where one can see that, after saturation, the oscillations occur at a somewhat higher frequency $\hat{\omega}_{n=1} \approx 0.0086$ ($3.16\,{\rm kHz}$), close to the slow magnetoacoustic plateau frequency $\omega_{\rm S0}$.
\end{enumerate}

\noindent The difference $0.0086 - 0.0072 = 0.0014$ between the frequencies in the exponentially growing and saturated phases is similar to the growth rate $\hat{\gamma}_{n=1} \approx 0.0015$. Indeed, in all our simulations (especially with large growth rates in Fig.~\ref{fig:06_flux0.50cut2_scan-diff}), we observe that
\begin{equation}
	\omega_{n=1} \approx \omega_{\rm S0} - \gamma_{n=1} \quad \text{for} \quad \gamma_{n=1} \geq 0.
	\label{eq:freq_shift_gamma}
\end{equation}

The right-hand side of Fig.~\ref{fig:03_flux0.50cut2_577-evol} shows the evolution of the entire mode structure (including $n=0$ and $|n|=1$) in the poloidal plane $(R,z)$ at $\varphi = 0$. The snapshot times are indicated by red and blue circles in panel (a). Panel set (e) shows how the peak of the normalized pressure $\beta$ oscillates back and forth, here in the major radial direction $R$ within the midplane ($z = z_0 \approx 0$). The fluctuating component $\delta\beta$ is shown in panel set (f). Comparison of the last snapshot of $\delta\beta$ in Fig.~\ref{fig:03_flux0.50cut2_577-evol}(f-8) with the snapshot of $2\delta \hat{B}_{\rm tor} \equiv 2\delta B_{\rm tor}/B_0$ in the additional panel (h-8) confirms that the pressure and magnetic perturbations satisfy the diamagnetic condition (\ref{eq:diamag_condition}) also in the saturated regime.

Panel set (g) of Fig.~\ref{fig:03_flux0.50cut2_577-evol} shows snapshots of the fluctuating toroidal current density $\delta J_{\rm tor}$. It has a central component near $R \approx R_0$ whose sign alternates with period $\tau_{n=1} = 2\tau_{\rm pulse}$, and two side lobes on the low- and high-field sides that flow in the opposite direction. The alternating $n=1$ component is nearly absent in panels (g-4) and (g-8), which are dominated by the axisymmetric component $[\delta J_{\rm tor}]_{n=0}$ whose structure is shown in the additional panel (i-8). Similarly to the diamagnetic condition (\ref{eq:diamag_condition}) for $\delta B_\parallel \approx \delta B_{\rm tor}$, we interpret the structure of $\delta J_{\rm tor}$ in Fig.~\ref{fig:03_flux0.50cut2_577-evol}(g) as a manifestation of the system trying to maintain a state of near-MHD-force-balance, with $\delta J_\parallel \approx \delta J_{\rm tor}$ reflecting changes in the poloidal magnetic field in response to the changed pressure field in Fig.~\ref{fig:03_flux0.50cut2_577-evol}(e,f). Indeed, magnetic Poincar\'{e} plots (not shown here) indicate small changes in the magnetic axis location (Shafranov shift) and the form of flux surfaces.

It should be noted that, unlike the $n=1$ Fourier component that oscillates once in two pulses, the $n=0$ component of the fluctuating fields (including $[\delta J_{\rm tor}]_{n=0}$ in Fig.~\ref{fig:03_flux0.50cut2_577-evol}(i-8)) oscillates once per pulse:
\begin{equation}
	\omega_{n=0} = \omega_{\rm pulse} = 2\omega_{n=1},
	\label{eq:freq_n01}
\end{equation}

\noindent because the perturbed field structures in Fig.~\ref{fig:03_flux0.50cut2_577-evol}(e,f,g) pass through the same $n=0$ state during each zero-crossing of the $n=1$ oscillations.

The waveforms of the minor radial, poloidal and toroidal components of the MHD velocity $\delta{\bm u}$ are shown on the left-hand side of Fig.~\ref{fig:04_flux0.50_sat-trace} for Fourier components $(m,n) = (0,0)$ and $(1,1)$ measured at the mode's peak radius $\psi = 0.05$. The right-hand side of Fig.~\ref{fig:04_flux0.50_sat-trace} compares the waveforms of the $(1,1)$ Fourier component of the electrostatic potential $\delta\Phi$, normalized pressure $\beta$, and toroidal MHD velocity $\delta u_{\rm tor}$ in more detail in the region $0.02 \leq \psi \approx 0.07$. Besides confirming Eq.~(\ref{eq:freq_n01}), the following observations can be made in Fig.~\ref{fig:04_flux0.50_sat-trace}:
\begin{enumerate}
	\item[(vi)]  The mode's poloidal phase (as determined by the arbitrarily chosen initial perturbation) remains unchanged for the first $5000$ Alfv\'{e}n times of the simulation, where either the cosine or the sine component of $n=1$ dominates, while the other is vanishingly small. This situation changes only in the late stages of the simulations, which we shall not discuss here and note only that the oscillations in the $(0,0)$ components of $\delta u_{\rm pol}$ and $\delta u_{\rm tor}$ in Fig.~\ref{fig:04_flux0.50_sat-trace}(b,c) eventually attain significant amplitudes.

	\item[(vii)] Fig.~\ref{fig:04_flux0.50_sat-trace}(j,k) shows that the normalized pressure $\delta\beta$ and the toroidal MHD velocity $[\delta u_{\rm tor}]_{1,1}$ perform effectively harmonic oscillations across the with of the mode. This indicates that the (toroidal) magnetoacoustic component of saturated SMAPs is robust and continues to behave largely like a linear wave. In contrast, the radial and poloidal components $[\delta u_{\rm rad}]_{1,1}$ in (c) and $[\delta u_{\rm pol}]_{1,1}$ in (f) have approximately harmonic form only around $\psi \approx 0.05...0.07$, whereas significant distortions are seen at smaller radii, as the waveform of $\delta\Phi_{1,1} \propto -i[\delta u_{\rm rad}]_{1,1}$ in Fig.~\ref{fig:04_flux0.50_sat-trace}(i) shows.
	
	\item[(viii)] Taking $[\delta u_{\rm tor}]_{1,1}^{\rm cos}$ in Fig.~\ref{fig:04_flux0.50_sat-trace}(f,k) as a reference, we see that $\delta\Phi_{1,1}^{\rm sin}$ and $\delta\beta^{\rm cos}$ in Fig.~\ref{fig:04_flux0.50_sat-trace}(g,i,j) lead by $\omega\Delta t_{\rm lead} \approx +\pi/2$ (more than the $+0.38\pi$ seen during exponential growth in Fig.~\ref{fig:02_flux0.50cut2_577-evol-exp}(a)). In Fig.~\ref{fig:04_flux0.50_sat-trace}(i), the distorted waveforms of $\delta\Phi_{1,1}^{\rm sin}(\psi < 0.05)$ closer to the center lead by an even larger phase shift. The relative phasing between $[\delta u_{\rm pol}]_{1,1}$ and $[\delta u_{\rm tor}]_{1,1}$ is also sensitive to the measurement location, but at the (apparently special) radius of $\psi = 0.05$ in Fig.~\ref{fig:04_flux0.50_sat-trace}(e,f), they are roughly out-of-phase ($\pm\pi$).
	
	\item[(ix)] The magnitude of the oscillations in the toroidal and poloidal MHD velocity components in Fig.~\ref{fig:04_flux0.50_sat-trace}(e,f) have a ratio of about $|[\delta u_{\rm tor}]_{1,1}|/|[\delta u_{\rm pol}]_{1,1}| \approx 20$, which shows the strong compressibility (weak Alfv\'{e}nicity) of the waves.
	
	\item[(x)] Consistently with Fig.~\ref{fig:04_flux0.50_sat-trace}(i), the snapshots of the electrostatic potential $\delta\Phi_{n=1}(R,z)$ in panel set (h) show that the wave phase flip between successive pulses is realized by a combination of radially outward-propagating ${\bm E}\times{\bm B}$ vortex patterns coming from the magnetic axis, and radially inward propagating patterns coming from the edge of the $q\sim 1$ region. The patterns converge and merge around $0.05 \lesssim \psi \lesssim 0.07$, as one can infer from the respective black contours drawn in Fig.~\ref{fig:04_flux0.50_sat-trace}(h-1) and (h-8). In snapshots of $\delta u_{\rm tor}$ (not shown), such patterns can be seen only at very small amplitudes, close to the signal's zero crossings. As described in item (iii) above, something similar was also seen during exponential growth in Fig.~\ref{fig:02_flux0.50cut2_577-evol-exp}(e,g), leading us to speculate that we are seeing a newly formed quasi-mode during each pulse.
\end{enumerate}

\noindent Implications of these and other observations that will be described in the following subsections will later be used to construct a physical picture of SMAPs in Section~\ref{sec:discussion}.

\subsection{Role of plasma compressibility}
\label{sec:results_gamma}

In order to confirm that the frequency of our core-localized quasi-mode is determined by the central plateau of the slow magnetoacoustic continuum (rather than being there by mere coincidence), we restarted the simulation of Fig.~\ref{fig:03_flux0.50cut2_577-evol} from $\hat{t} = 6857$ ($\approx 3\,{\rm ms}$) and continued it for another $2000$ Alfv\'{e}n times ($\approx 1\,{\rm ms}$) with three different values of the specific heat ratio $\Gamma$, which controls plasma compressibility. The results of this numerical experiment are summarized in Fig.~\ref{fig:05_flux0.50cut2_scan-G}.

When continuing the simulation with the default value $\Gamma = 5/3$ (red dashed), the pulsations of $W_{n=1}(t)$ in Fig.~\ref{fig:05_flux0.50cut2_scan-G}(a) continue at the same pace and the quasi-mode's oscillation frequency $\hat{\omega}_{n=1}(\Gamma = 5/3) = 0.0084$ ($3.16\,{\rm kHz}$) in Fig.~\ref{fig:05_flux0.50cut2_scan-G}(c) remains near the slow magnetoacoustic plateau frequency $\hat{\omega}_{\rm S0}(\Gamma = 5/3) = 0.0085$ ($3\,{\rm kHz}$). When we increase the specific heat ratio to $\Gamma = 3$ (magenta), the pulsations in Fig.~\ref{fig:05_flux0.50cut2_scan-G}(a) speed up while reducing their amplitude. The measured frequency $\hat{\omega}_{n=1}(\Gamma = 3) = 0.0108$ ($3.97\,{\rm kHz}$) matches well the increased value of $\hat{\omega}_{\rm S0}$ in Fig.~\ref{fig:05_flux0.50cut2_scan-G}(d). When we use a lower $\Gamma = 1$ (light blue), the pulsations in Fig.~\ref{fig:05_flux0.50cut2_scan-G}(a) slow down while elevating their amplitude, and the measured frequency $\hat{\omega}_{n=1}(\Gamma = 1) = 0.0069$ ($2.54\,{\rm kHz}$) matches well the reduced value of $\hat{\omega}_{\rm S0}$ in Fig.~\ref{fig:05_flux0.50cut2_scan-G}(b). This confirms that it is the value of the slow magnetoacoustic continuum plateau frequency in the region of nearly flat safety factor $q \gtrsim 1$ that sets the pace of the pulsations.

\begin{figure}
	[tb!]\vspace{-0.3cm}
	\centering
	\includegraphics[width=0.48\textwidth]{\figures/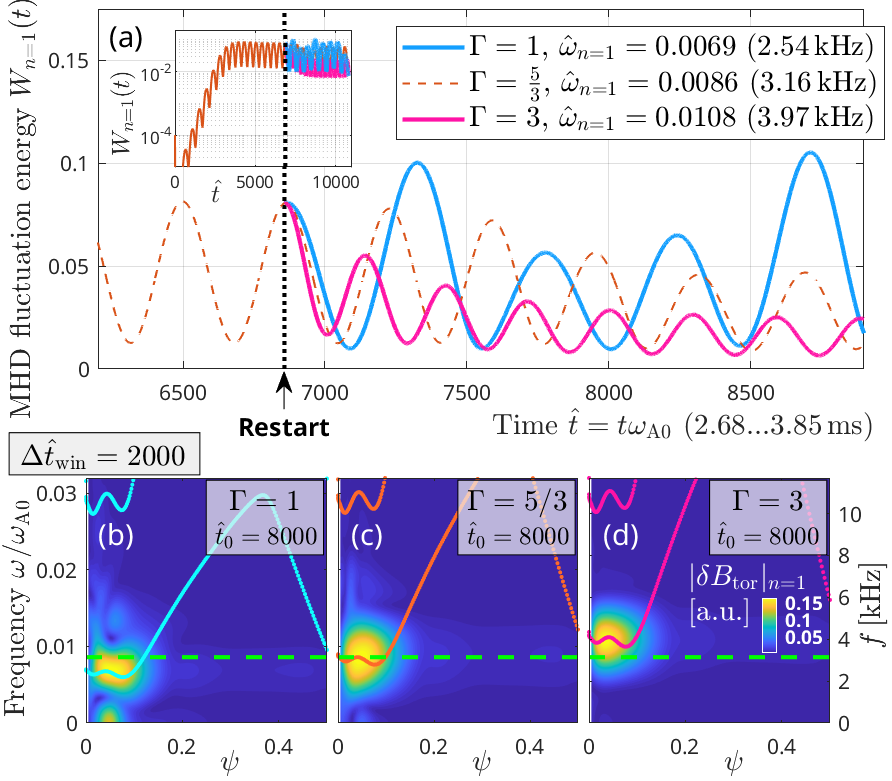}\vspace{-0.3cm}
	\caption{Continuation of the simulation of Fig.~\protect\ref{fig:03_flux0.50cut2_577-evol} with three different values of the specific heat ratio: reduced $\Gamma = 1$ (blue), default $\Gamma = 5/3$ (brown), increased $\Gamma = 3$ (magenta). The simulation was restarted at $\hat{t} = 6857$ ($\approx 3\,{\rm ms}$) and panel (a) shows the time traces of the magnetic fluctuation energy $W_{n=1}(t)$ for the subsequent $2000$ Alfv\'{e}n times ($\approx 1\,{\rm ms}$). Panels (b)--(d) show the respective spectrograms. The horizontal dashed green line indicates the wave frequency $\hat{\omega}_{n=1} = 0.0086$ ($3.16\,{\rm kHz}$) in the default case (c) for comparison. The colors of the overlaid MHD continua in (b)--(d) correspond to those used in panel (a) for different values of $\Gamma$.}
	\label{fig:05_flux0.50cut2_scan-G}\vspace{-0.3cm}
\end{figure}

\begin{figure*}
	[tb!]\vspace{-1.7cm}
	\centering
	\includegraphics[width=0.96\textwidth]{\figures/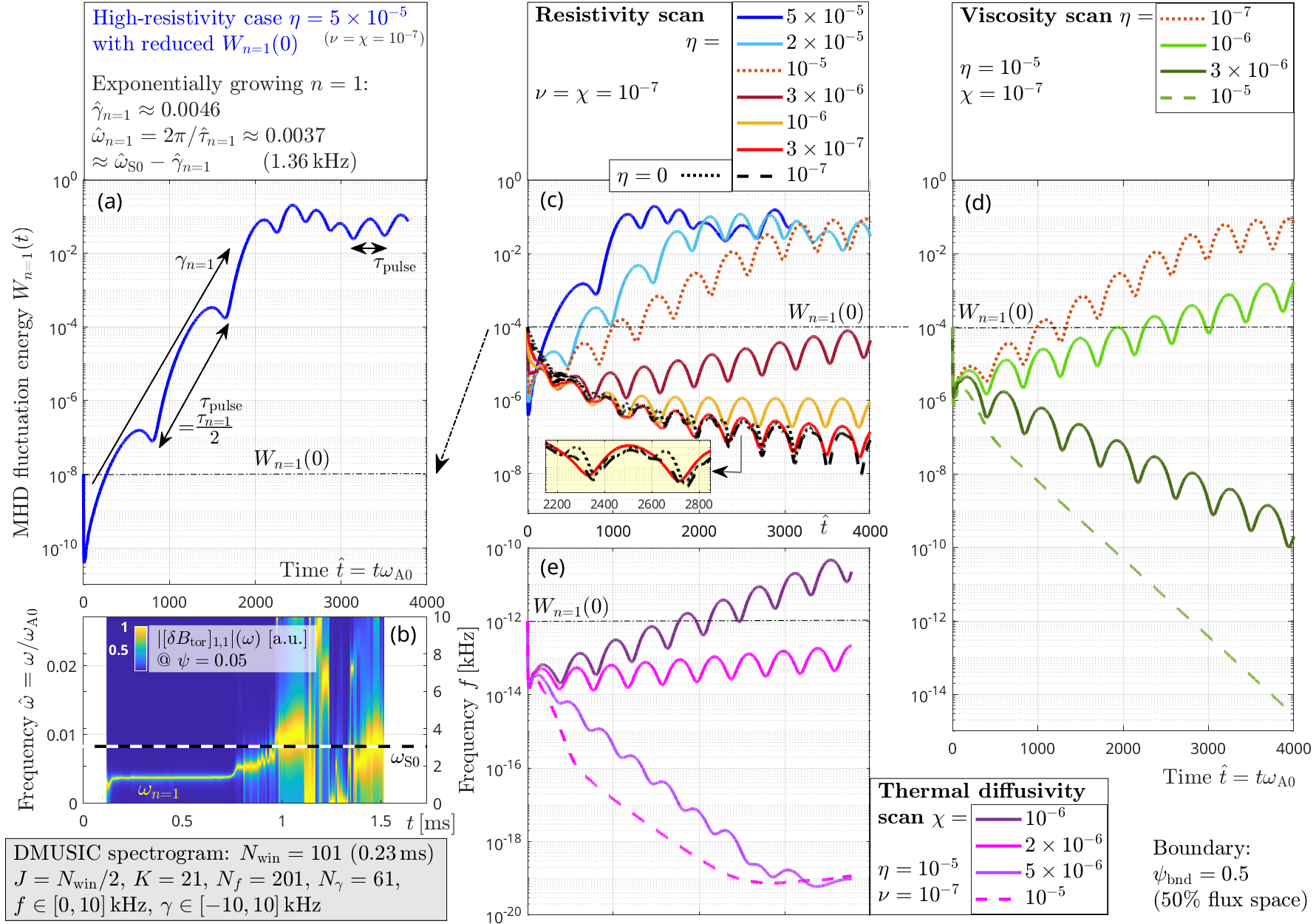}\vspace{-0.1cm}
	\caption{Influence of the diffusion coefficients $\eta$, $\nu$, $\chi$ on the stability (and existence) of slow magnetoacoustic pulsations (SMAPs) in simulations of the inner $50\%$ of flux space. Panels (a,b) show results for the case with the highest resistivity ($\eta = 5\times 10^{-5}$) and lowest viscosity and diffusivity ($\nu = \chi = 10^{-7}$) of our scan. The high growth rate $\hat{\gamma}_{n=1} = 0.0046 \approx \hat{\omega}_{\rm S}/2$ in this case demonstrates clearly the resulting down-shift $\hat{\omega}_{n=1} \approx \hat{\omega}_{\rm S0} - \hat{\gamma}_{n=1}$ of the wave frequency during the exponential growth phase ($\hat{t} < 2000$). A low initial perturbation amplitude giving $W_{n=1}(0) \approx 10^{-8}$ was used to allow the quasi-mode to grow through a full oscillation period $\tau_{n=1} = 2\tau_{\rm pulse}$. The {\tt DMUSIC} spectrogram $|\delta B_{\rm tor}|_{n=1}(\omega)$ of the toroidal magnetic fluctuations evaluated at $\psi = 0.05$ is plotted in panel (b) and shows how the quasi-mode frequency $\omega_{n=1}$ rises during saturation ($\hat{t} \approx 2000$) and subsequently hovers near the central plateau frequency $\hat{\omega}_{\rm S0} \approx 0.0083$ ($3\,{\rm kHz}$) of the slow magnetoacoustic continuum's $(m,n)=(1,1)$ branch. The {\tt DMUSIC} time window comprised $N_{\rm win} = 101$ diagnostic samples, which corresponds to $\Delta\hat{t}_{\rm win} \approx 540$ ($0.23\,{\rm ms}$). For explanations regarding the other {\tt DMUSIC} parameters that are shown in the gray text box, see Appendix E.2 of Ref.~\protect\cite{Bierwage21} and the original paper by Li {\it et al}.\ \cite{LiY98}. Panel (c) shows results of a resistivity scan over the range $0 \leq \eta \leq 5\times 10^{-5}$ for fixed $\nu = \chi = 10^{-7}$. Since $\eta$ is destabilizing, the legend for (c) shows high to low $\eta$ values from top to bottom. The blue curve is the same case as in panel (a), but with higher initial perturbation amplitude $W_{n=1}(0) = 10^{-4}$. Panel (d) shows results of a viscosity scan over the range $10^{-7} \leq \nu \leq 10^{-5}$ for fixed high resistivity $\eta = 10^{-5}$ and low diffusivity $\chi = 10^{-7}$. The inset at the bottom of (c) shows an enlarged view of the pulse shape at low $\eta \leq 3\times 10^{-7}$. (See Fig.~\protect\ref{fig:a01_flux0.50_eta0_convergence} for a numerical convergence test for $\eta = 0$.) Panel (e) shows results of a diffusivity scan over the range $10^{-6} \leq \chi \leq 10^{-5}$ for fixed high resistivity $\eta = 10^{-5}$ and low viscosity $\nu = 10^{-7}$, beginning from a very small $W_{n=1}(0) = 10^{-12}$ to demonstrating that the SMAPs are independent of the initial perturbation amplitude. The dotted brown curves in panels (c) and (d) are our standard case with high $\eta = 10^{-5}$ and low $\nu= \chi= 10^{-7}$, which was examined in more detail in Figs.~\protect\ref{fig:03_flux0.50cut2_577-evol}--\protect\ref{fig:05_flux0.50cut2_scan-G} above.}
	\label{fig:06_flux0.50cut2_scan-diff}\vspace{-0.15cm}
\end{figure*}

\begin{figure}
	[tb!]\vspace{-2.5cm}
	\centering
	\includegraphics[width=0.48\textwidth]{\figures/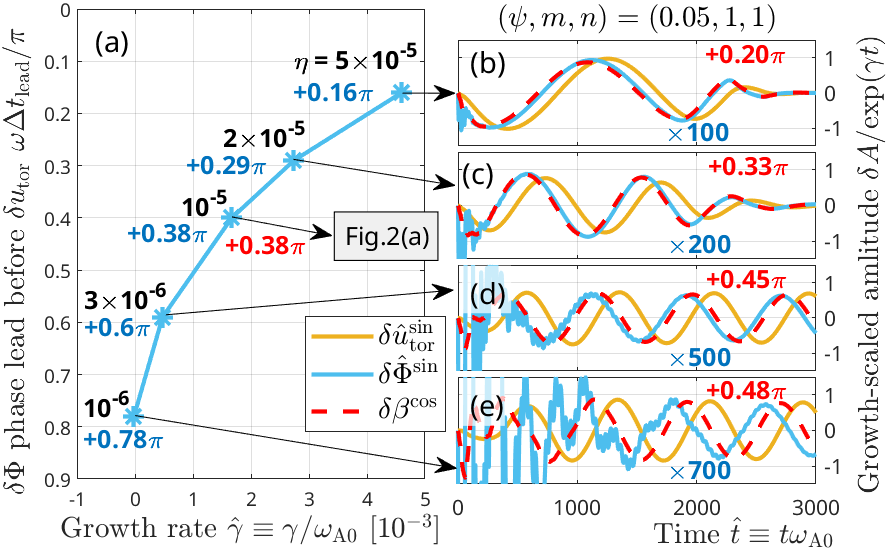} \\
	\includegraphics[width=0.48\textwidth]{\figures/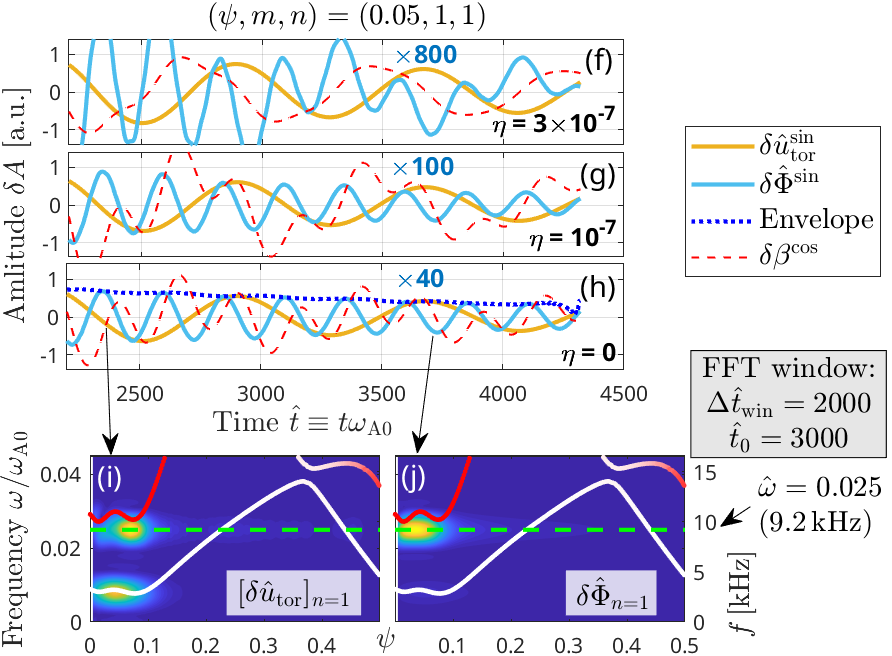}\vspace{-0.1cm}
	\caption{Resistivity-dependence of the phase lead of $\delta\Phi$ (blue) and $\delta\beta$ (red) relative to $\delta u_{\rm tor}$, and emergence of higher-frequency pulsations near the Alfv\'{e}nic continuum plateau. Panel (a) shows an overview of the measured values of the phase lead $\omega\Delta t_{\rm lead}$ (vertical axis) as a function of the growth rate $\gamma(\eta)$ (horizontal axis) and resistivity $\eta$ (black labels). The blue and red labels are the numerical values of $\omega\Delta t_{\rm lead}$ by which $\delta\Phi^{\rm sin}_{1,1}$ and $\beta^{\rm cos}_{1,1}$ lead ahead of $[\delta u_{\rm tor}]_{1,1}^{\rm sin}$. For the marginally stable and unstable cases $10^{-6} \leq \eta \leq 5\times 10^{-5}$, the raw data are shown in panels (b)--(e) in the form of time traces of the $(m,n)=(1,1)$ Fourier components all three signals measured at $\psi = 0.05$. Here, the exponential growth was eliminated by applying a factor $\exp(-\gamma t)$. Moreover, the $\delta\Phi$ signal ($\propto {\bm E}\times{\bm B}$ flow) was scaled up by the numerical factors shown in each panel in order to be visible on the same scale as $\delta u_{\rm tor}$. A larger scaling factor is interpreted as indicating a higher compressibility and weaker Alfv\'{e}nicity. The amplitude of $\delta\beta$ was also rescaled for ease of comparison.  Panels (f)--(h) show the time traces for the stable cases $0\leq \eta \leq 3\times 10^{-7}$, where higher-frequency Alfv\'{e}nic oscillations emerge and eventually dominate over SMAPs in the $\delta\Phi$ signal that represents the electromagnetic component of the fluctuations. Panels (i) and (j) show the respective spectrograms of $\delta u_{\rm tor}$ and $\delta\Phi$ for the zero-resistivity case of (h). Acoustic (white) and Alfv\'{e}nic (red) continua are shown for comparison.}
	\label{fig:07_flux0.50_phase-eta}\vspace{-0.2cm}
\end{figure}

\begin{figure}
	[tb!]\vspace{-2.5cm}
	\centering
	\includegraphics[width=0.48\textwidth]{\figures/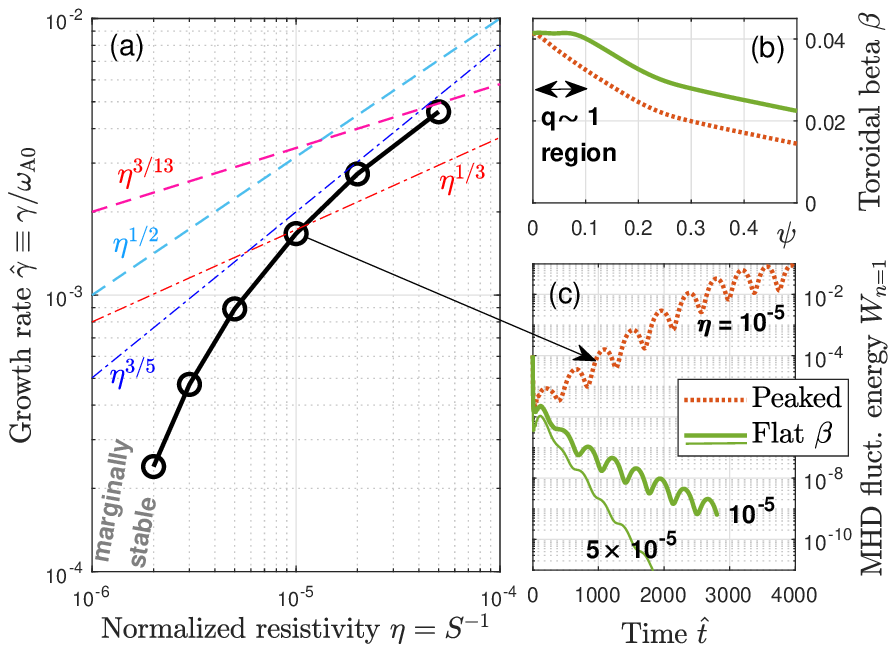}\vspace{-0.2cm}
	\caption{Dependence of SMAP growth on resistivity and the central pressure gradient. Panel (a) shows on a log-log scale the resistivity-dependence of the growth rate $\hat{\gamma}$ using the 4 unstable cases of Fig.~\protect\ref{fig:06_flux0.50cut2_scan-diff}(c) plus two intermediate data points with $\eta = 2\times 10^{-6}$ and $5\times 10^{-6}$. For comparison, dash-dotted and dashed lines indicate the scalings $\eta^{3/13}$, $\eta^{1/3}$, $\eta^{1/2}$, and $\eta^{3/5}$ as labeled. Panel (b) shows as a dotted brown curve our default normalized pressure profile $\beta_{\rm ref}(\psi) = 2\mu_0 P_{\rm ref}(\psi)/B_0^2$ and a modified profile $\beta_{\rm flat}(\psi)$ (solid green line) that has a negligible gradient in the region $\psi \lesssim 0.1$, where $q\sim 1$ in Fig.~\protect\ref{fig:01_setup}(c). These are the same pressure profiles as in Fig.~\protect\ref{fig:01_setup}(b), but shown here only in the reduced poloidal flux domain $0 \leq \psi \leq 0.5$, where the present simulations were performed. Panel (c) shows the time traces of the MHD fluctuation energy $W_{n=1}(t)$ obtained with the two pressure profiles of panel (b) and resistivity $\eta = 10^{-5}$, as well as the result obtained with the centrally flat $\beta$ profile and a higher resistivity $\eta = 5\times 10^{-5}$.}
	\label{fig:08_flux0.50_growth-eta}\vspace{-0.2cm}
\end{figure}

\subsection{Role of non-ideal effects (diffusion)}
\label{sec:results_diff}

Fig.~\ref{fig:06_flux0.50cut2_scan-diff} summarizes results of scanning the values of the diffusion coefficients ($\eta$, $\nu$, $\chi$) in simulations of the reduced domain ($\psi_{\rm bnd} = 0.5$). Fig.~\ref{fig:06_flux0.50cut2_scan-diff}(c) shows time traces of the MHD fluctuation energy $W_{n=1}(t)$ for resistivities in the range $0 \leq \eta \leq 5\times 10^{-5}$ while fixing the viscosity and diffusivity at a low value $\nu=\chi=10^{-7}$. Figs.~\ref{fig:06_flux0.50cut2_scan-diff}(d) and \ref{fig:06_flux0.50cut2_scan-diff}(e) show $W_{n=1}(t)$ for viscosities in the range $10^{-5} \leq \nu \leq 10^{-7}$ and diffusivities in the range $10^{-5} \leq \chi \leq 10^{-6}$ while keeping $\eta = 10^{-5}$ relatively large and $\chi=10^{-7}$ or $\nu=10^{-7}$ small. The simulations were also initialized with several different perturbation amplitudes, giving initial $W_{n=1}(0) = 10^{-8}$ in (a), $W_{n=1}(0) = 10^{-4}$ in (c,d), and $W_{n=1}(0) = 10^{-12}$ in (e). These result tell us the following about SMAPs:
\begin{itemize}
	\item SMAPs have no excitation threshold in the initial perturbation amplitude, so $W_{n=1}(0)$ can be chosen arbitrarily.
	
	\item The pulsations vanish if the ion viscosity becomes large enough to cause a collapse of the associated ${\bm E}\times{\bm B}$ vortices --- which happens in Fig.~\ref{fig:06_flux0.50cut2_scan-diff}(d) for $\nu > 3\times 10^{-5}$ --- or if the thermal diffusivity becomes large enough to suppress thermal misbalance --- which happens in Fig.~\ref{fig:06_flux0.50cut2_scan-diff}(e) for $\chi > 5\times 10^{-6}$.
	
	\item Resistivity $\eta$ strongly affects the growth rate in Fig.~\ref{fig:06_flux0.50cut2_scan-diff}(a,c), but not the log-scale magnitude of SMAPs nor their existence. Indeed, the overlapping red and black curves show that SMAPs become independent of resistivity for $\eta \lesssim 3\times 10^{-7}$.\footnote{This also rules out the possibility that the loop voltage $E_{\parallel,{\rm src}} = -\eta_{\rm e} J_{\rm ref}$ in Ohm's law (\protect\ref{eq:model_cc_efield}) is required for SMAP existence (e.g., by counteracting oscillations of the magnetic axis or $q$ profile).}
\end{itemize}

\noindent In the remainder of this section, we discuss in more detail observations that can be made in the resistivity scan.

In our reference case with $\eta = 10^{-5}$ and $\nu=\chi = 10^{-7}$ --- whose $W_{n=1}(t)$ is plotted as a dotted brown curve in panels (c) and (d) of Fig.~\ref{fig:06_flux0.50cut2_scan-diff}, and which we already inspected in detail in the previous subsections \ref{sec:results_0.5} and \ref{sec:results_gamma} --- the growth rate $\gamma_{n=1}/\omega_{n=1} \sim 0.21$ can be said to be in the strongly driven regime. For the largest resistivity in our scan, $\eta \gtrsim 5\times 10^{-5}$ (blue), the growth rate even exceeds the oscillation frequency: $\gamma_{n=1}/\omega_{n=1} \sim 1.24$. The consequence of this can be seen in panels (a) and (b) of Fig.~\ref{fig:06_flux0.50cut2_scan-diff}, where the pulsation period $\tau_{\rm pulse} = \tau_{n=1}/2$ is significantly lengthened during the phase of exponential growth ($\hat{t} < 2000$) and the corresponding spectral peak appears at a significantly reduced frequency $\hat{\omega}_{n=1} = 2\pi/\hat{\tau}_{n=1} = 0.0037$ ($1.36\,{\rm kHz}$). This case confirms again an observation we made in Section~\ref{sec:results_0.5} above, where we noted in Eq.~(\ref{eq:freq_shift_gamma}) that the frequency of the growing mode is down-shifted from the slow magnetoacoustic continuum plateau by the growth rate; here: $\hat{\omega}_{n=1} \approx \hat{\omega}_{\rm S0} - \hat{\gamma}_{n=1} \approx 0.0083 + 0.0046 = 0.0037$. After saturation ($t > 2000$) we have $\omega_{n=1} \approx \omega_{\rm S0}$, and the transition has the appearance of an upward chirp in the high-resolution {\tt DMUSIC} \cite{LiY98} spectrogram in Fig.~\ref{fig:06_flux0.50cut2_scan-diff}(b).

Fig.~\ref{fig:07_flux0.50_phase-eta}(a)--(e) shows how the electric resistivity $\eta$ affects the phase lead of $\delta\Phi$ (blue) and $\delta\beta$ (red) relative to $\delta u_{\rm tor}$ in SMAPs during the exponentially growing phase. In Fig.~\ref{fig:02_flux0.50cut2_577-evol-exp}(a) above, both $\delta\Phi$ and $\delta\beta$ were leading ahead of $\delta u_{\rm tor}$ by approximately $\omega\Delta t_{\rm lead} \approx +0.38\pi$ for $\eta = 10^{-5}$ and $\nu=\chi = 10^{-7}$. The $\eta$-scan in Fig.~\ref{fig:07_flux0.50_phase-eta} shows that the phase lead of $\delta\Phi$ decreases towards zero for increasing resistivity $\eta$ and growth $\gamma(\eta)$, and increases towards $+\pi$ (out-of-phase) for decreasing resistivity $\eta$ and $\gamma(\eta)$. At the same time, the ratio $|\delta u_{\rm tor}|/|\delta\Phi|$ increases with decreasing resistivity and growth rate as can be inferred from the numerical scaling factors that are printed blue at the bottom of panels (b)--(e) of Fig.~\ref{fig:07_flux0.50_phase-eta}, indicating that the electromagnetic (Alfv\'{e}nic) component weakens as $\gamma(\eta)$ decreases. Moreover, one can see in panels (b)--(e) that it takes longer for the erratic transients in $\delta\Phi(t)$ to decay, while the acoustic wave comprising $\delta\beta$ and $\delta u_{\rm tor}$ is established almost instantaneously. The fact that the SMAPs in Fig.~\ref{fig:06_flux0.50cut2_scan-diff}(c) begin to grow only after the waveform of $\delta\Phi(t)$ has been established constitutes a strong evidence suggesting that the electromagnetic component represented by $\delta\Phi$ ($\propto {\bm E}\times{\bm B}$ flow) as well as the non-trivial relative phasing between $\delta\Phi$, $\delta\delta$ and $\delta u_{\rm tor}$ plays a crucial role for destabilizing SMAPs.

The phase lead of $\delta\beta$ (red numbers in Fig.~\ref{fig:07_flux0.50_phase-eta}) ahead of $\delta u_{\rm tor}$ varies between values close to $0$ for rapid growth and $\pi/2$ near marginal stability, which makes physical sense for a slow magnetoacoustic wave, as we will discuss later in Section~\ref{sec:discussion}.

While the acoustic branch dominates near marginal stability, the situation changes when the resistivity is reduced further, as one can see in the lower part of Fig~\ref{fig:07_flux0.50_phase-eta}, where panels (f)--(j) show results for $0 \leq \eta \lesssim 3\times 10^{-7}$. Here, the SMAP signal in $\delta\Phi$ becomes increasingly obscured by the emergence of higher-frequency pulsations. These were also visible in the time traces of the volume-integrated MHD fluctuation energy $W_{n=1}(t)$ in the enlarged inset at the bottom of Fig.~\ref{fig:06_flux0.50cut2_scan-diff}(c), especially for $\eta = 0$ (dotted black). The spectrograms in Fig.~\ref{fig:07_flux0.50_phase-eta}(i,j) show that this frequency $\omega \approx 0.025\omega_{\rm A0} \approx 2\pi\times 9\,{\rm kHz}$ is located just below the central plateau of the Alfv\'{e}nic continuum branch. One can see that this branch dominates the electromagnetic signal represented by $\delta\Phi$ when $\eta \lesssim 3\times 10^{-7}$, while the acoustic branch remains strong in $\delta u_{\rm tor}$. For $\eta = 0$, Fig.~\ref{fig:07_flux0.50_phase-eta}(h) shows that both signals decay at the same rate (presumably via viscous damping). The increasing strength of the electromagnetic component can also be quantitatively inferred from the decrease of the numerical scaling factors printed in blue at the top of panels (f)--(h) in Fig.~\ref{fig:07_flux0.50_phase-eta}.

Interestingly, the Alfv\'{e}nic pulsations remain active when we suppress SMAPs by increasing the thermal diffusivity as $\chi = 10^{-7}\rightarrow 10^{-5}$ (for fixed $\eta = 0$, $\nu = 10^{-7}$). This is shown in Fig.~\ref{fig:13_flux0.50_Alfvenic-pulse} of Section~\ref{sec:discussion}, which contains a brief discussion and outlook of this topic.
 
Finally, Fig.~\ref{fig:08_flux0.50_growth-eta}(a) summarizes the measured growth rates of our resistivity scan on a logarithmic scale. For comparison, some scaling laws that can be found in the literature are also shown: $\eta^{1/3}$ is found in studies of pressure-gradient-driven resistive interchange (or ballooning) and turbulence \cite{Furth63, Carreras87}. Depending on whether infernal modes \cite{Charlton89} are very close or further away from their ideal threshold, their scaling is predicted to lie between $\eta^{3/13}$ and the resistive tearing scaling $\eta^{3/5}$. Lastly, we show the scaling $\eta^{1/2}$ that has been reported in association with resistive ion acoustic waves \cite{Fernandez08}. Of course, such scaling laws are obtained mathematically only under very specific conditions and are altered by other nonideal effects, such as viscosity \cite{Furth63}. Moreover, those scaling laws are valid for purely growing MHD eigenmodes that do not have an instability threshold in $\eta$ (unlike SMAPs). In contrast, Fig.~\ref{fig:08_flux0.50_growth-eta}(a) shows that our SMAPs do not exhibit any particular scaling. The curve $\gamma(\eta)$ looks somewhat tangent to the $\eta^{3/5}$ and $\eta^{1/2}$ lines at high resistivity $10^{-5} < \eta < 10^{-4}$, but this may be coincidental and without relevance. Towards lower resistivity, the $\gamma(\eta)$ curve steepens before terminating at the instability threshold near $\eta \sim 10^{-6}$.

\subsection{Role of the pressure gradient}
\label{sec:results_press}

The instability threshold depends on the pressure gradient in the central $q\sim 1$ region, which seems to be the source of free energy that drives the instabilities found in this work. To demonstrate this, we have prepared another equilibrium with a flat central pressure that was shown in Fig.~\ref{fig:01_setup}(b) and is plotted again in Fig.~\ref{fig:08_flux0.50_growth-eta}(b) in the reduced simulation domain $0 \leq \psi \leq 0.5$. The time traces of the MHD fluctuation energy $W_{n=1}(t)$ obtained with centrally peaked and flat $\beta$ profiles with $\eta = 10^{-5}$ are compared in Fig.~\ref{fig:08_flux0.50_growth-eta}(c). One can see that the pulsations persist but become stable in the absence of a pressure gradient. (The same is found for Alfv\'{e}nic pulsations in Fig.~\ref{fig:13_flux0.50_Alfvenic-pulse} that will be discussed in Section~\ref{sec:discussion}.)

Interestingly, Fig.~\ref{fig:08_flux0.50_growth-eta}(c) also shows that raising the resistivity to $\eta = 5\times 10^{-5}$ further stabilizes the pulsations in the case with centrally flat pressure. This trend is opposite to what was seen in Figs.~\ref{fig:06_flux0.50cut2_scan-diff}(c) and \ref{fig:08_flux0.50_growth-eta}(a) with the centrally peaked pressure profile, for which $\eta$ had a destabilizing influence.

\begin{figure}
	[tb!]\vspace{-1.3cm}
	\centering
	\includegraphics[width=0.48\textwidth]{\figures/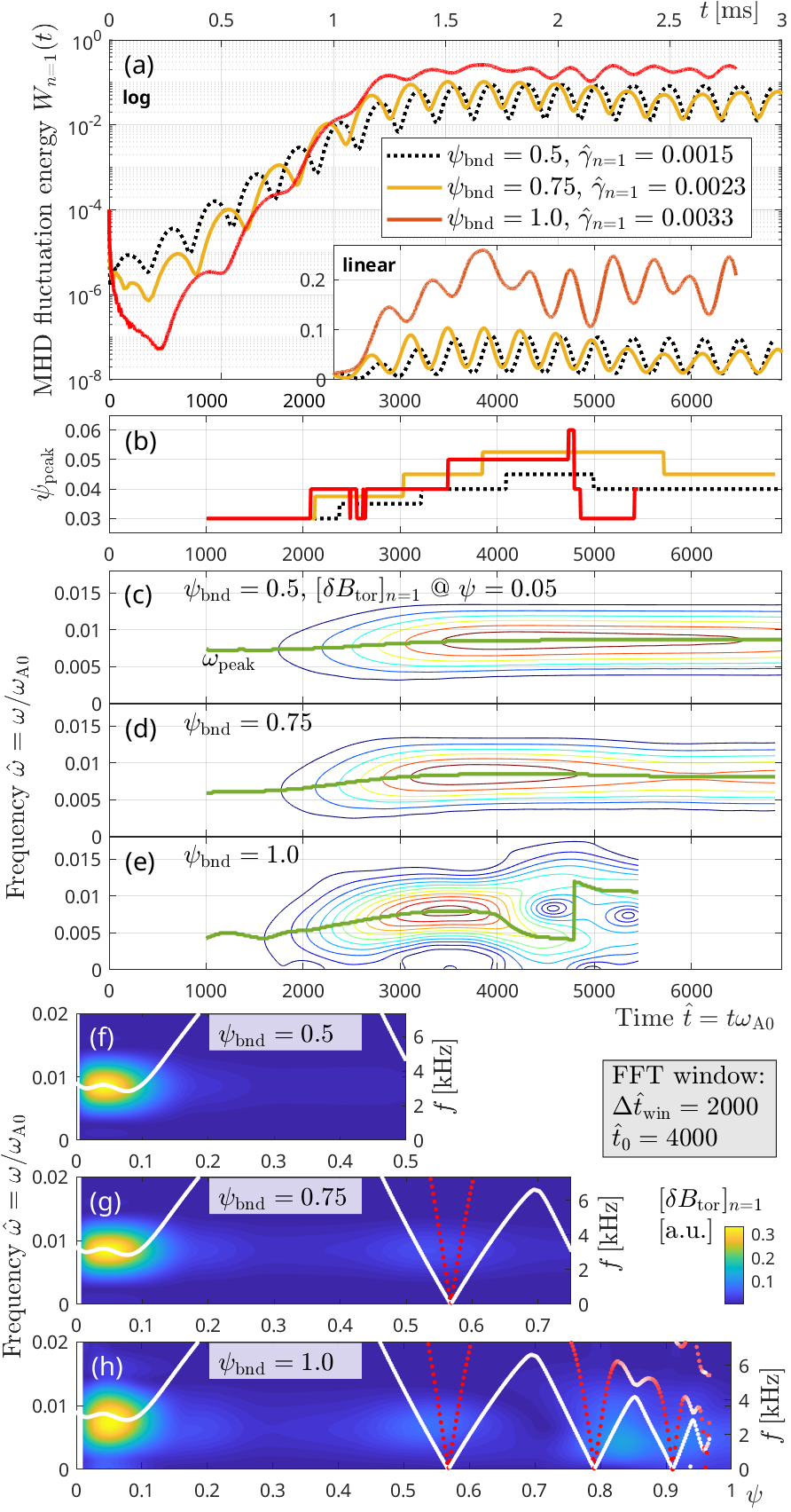}\vspace{-0.3cm}
	\caption{Comparison between simulations with different domain sizes, where the non-slip boundary condition is applied at $\psi_{\rm bnd} = 0.5$ (just inside $q=2$), $0.75$ (just inside $q=3$) or $1.0$ (just inside the last closed flux surface). Panels (a) and (b) show time traces of the respective MHD fluctuation energy $W_{n=1}(t)$ and mode peak location $\psi_{\rm peak}(t)$. Panels (c)--(e) show the temporal evolution of the Fourier spectrum computed from $[\delta B_{\rm tor}]_{n=1}$ at $\psi = 0.05$ with a time window of size $\Delta\hat{t}_{\rm win} = 2000$. The bold green line highlights the dominant frequency $\omega_{\rm peak}$. Panels (f)--(h) show global spectrograms $[\delta B_{\rm tor}]_{n=1}(\omega,\psi)$ for a time window centered around $\hat{t} = 4000$ ($1.73\,{\rm ms}$).}
	\label{fig:09_flux0.5-0.75-1.0_evol}\vspace{-0.3cm}
\end{figure}

\begin{figure*}
	[tb!]\vspace{-2.9cm}
	\centering
	\includegraphics[width=0.96\textwidth]{\figures/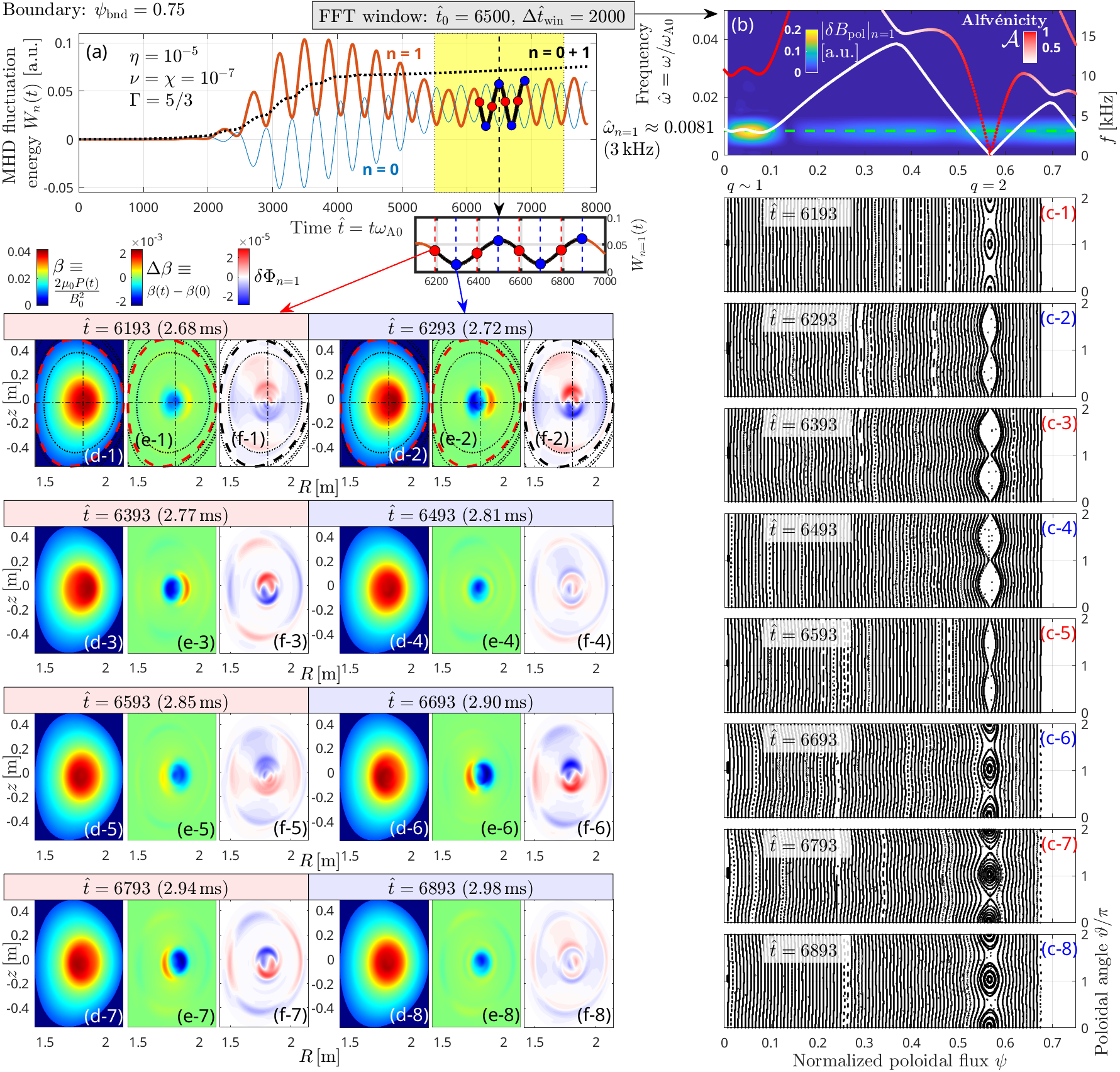}\vspace{-0.25cm}
	\caption{Evolution of slow magnetoacoustic pulsations (SMAPs) in a simulation covering the inner $75\%$ of the poloidal flux space of Fig.~\protect\ref{fig:01_setup}; that is, $0 \leq \psi \leq \psi_{\rm bnd} = 0.75$. Our standard values are used for diffusion and compressibility parameters ($\eta=10^{-5}$, $\nu = \chi = 10^{-7}$, $\Gamma = 5/3$). Panel (a) shows the evolution of the total MHD fluctuation energy $W_{n}(t)$ for $n=0$ (blue), $n=1$ (brown), and their sum (dotted black). The yellow shading indicates the FFT window $\hat{t}_0 \pm \Delta\hat{t}_{\rm win}/2$, where the spectrogram $|[\delta B_{\rm pol}]_{n~1}|(\omega,\psi)$ in panel (b) was computed. Overall, the contents of this figure are similar to those of Fig.~\protect\ref{fig:03_flux0.50cut2_577-evol}, except that here the emphasis is on the tearing-parity components, so the spectrogram in panel (b) was computed using the poloidal magnetic fluctuations $\delta B_{\rm pol}$ instead of the toroidal ones $\delta B_{\rm tor}$, panel set (f) shows snapshots of the electrostatic potential $\delta\Phi$ instead of the current density $\delta J_{\rm tor}$, and a Poincar\'{e} plot was made for each snapshot. The horizontal $\psi$-axis of the Poincar\'{e} plots in panel set (c) is aligned with the spectrogram in panel (b). The times of odd (red) and even (blue) snapshots are indicated by circles in panel (a). The ${\bm B}$ field lines were traced starting from $\vartheta=\varphi=0$, which coincides with the X-points in snapshots (c-2)--(c-5), so the island interiors were not sampled and thus appear empty there. The location of the $q=2$ surface can be inferred from the zeros of the MHD continua in panel (b), the island location in panel set (c), and a dotted contour line in panel sets (d)--(f). The horizontal and vertical dash dotted black lines in (d)--(f) intersect at the magnetic axis $(R_0,z_0)$.}\vspace{-0.3cm}
	\label{fig:10_flux0.75cut3_577-evol}%
\end{figure*}

\begin{figure*}
	[tb!]\vspace{-2.55cm}
	\centering
	\includegraphics[width=0.96\textwidth]{\figures/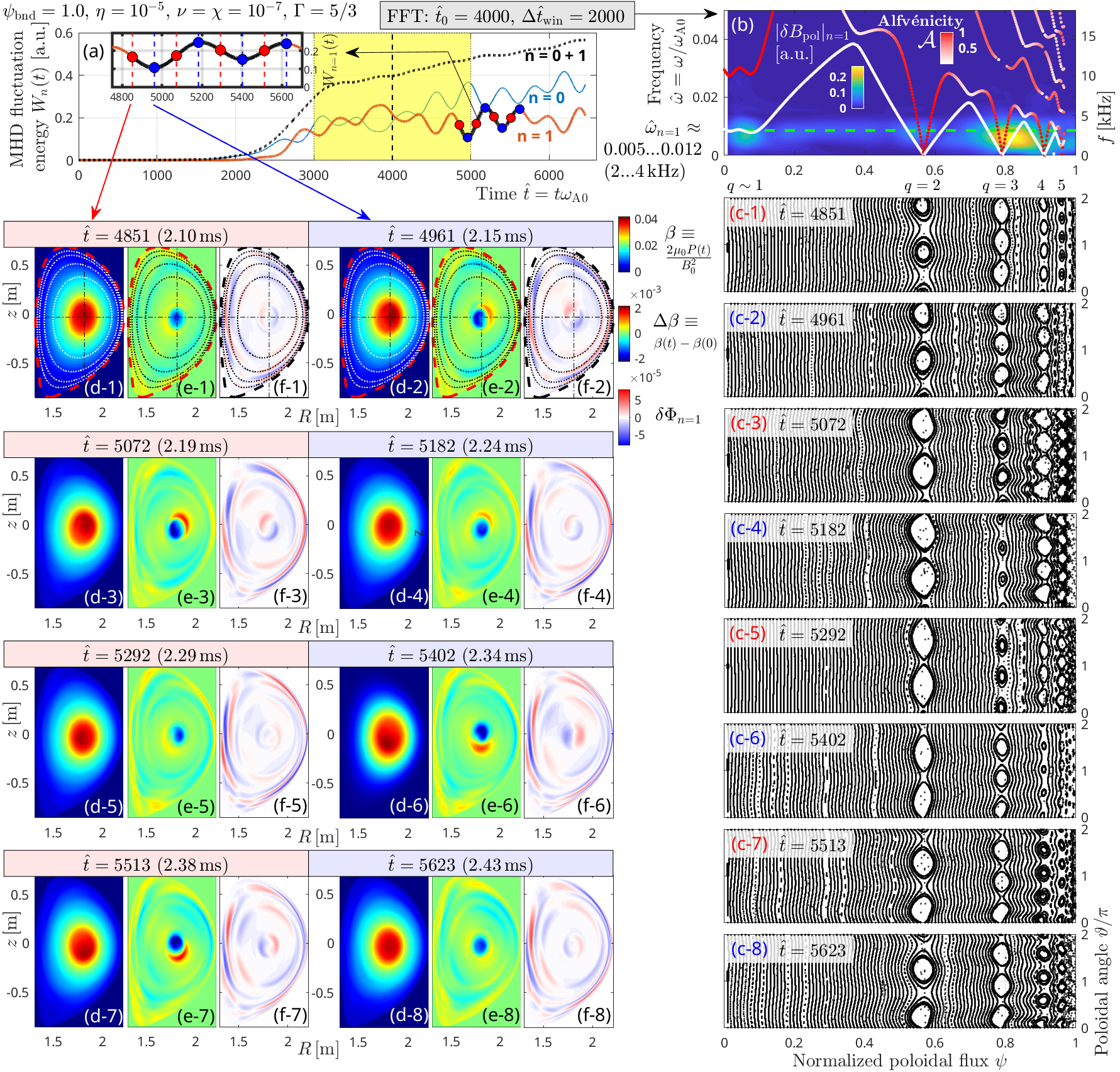}\vspace{-0.3cm}
	\caption{Evolution of SMAPs in a simulation covering the entire plasma within the last-closed flux surface of Fig.~\protect\ref{fig:01_setup}; that is, $0 \leq \psi \leq \psi_{\rm bnd} = 1.0$. Our standard values are used for diffusion and compressibility parameters ($\eta=10^{-5}$, $\nu = \chi = 10^{-7}$, $\Gamma = 5/3$). Arranged as Fig.~\protect\ref{fig:10_flux0.75cut3_577-evol}. The locations of the $q=2,3,4,5$ surfaces can be inferred from the zeros of the MHD continua in panel (b), the island locations in panel set (c), and the dotted contour lines in panel sets (d)--(f). The MHD continua in panel (b) were truncated at $\psi \approx 0.97$ due to poor resolution.}\vspace{-0.4cm}
	\label{fig:11_flux1.0_577-evol}%
\end{figure*}

\subsection{Interactions with $q \geq 2$ rational surfaces}
\label{sec:results_res}

Using our standard case with high $\eta = 10^{-5}$ and low $\nu = \chi = 10^{-7}$, we now proceed to simulations with non-slip boundaries placed at $\psi_{\rm bnd} = 0.75$, just outside the $q=2$ surface, and $\psi_{\rm bnd} = 1.0$ near the last closed flux surface. First, we compare in Fig.~\ref{fig:09_flux0.5-0.75-1.0_evol} the respective time traces of the MHD fluctuation energy $W_{n=1}(t)$ and spectrograms of the toroidal magnetic fluctuations $\delta B_{\rm tor}$ in simulations with the three domain sized $\psi_{\rm bnd} = 0.5,\,0.75,\,1.0$ (or, equivalently: $50\%,\, 75\%,\, 100\%$).

Fig.~\ref{fig:09_flux0.5-0.75-1.0_evol}(a) shows that the exponential growth rate increases significantly with increasing domain size from $\hat{\gamma}_{n=1}(50\%) = 0.0015$ (black dotted) via $\hat{\gamma}_{n=1}(75\%) = 0.0023$ (orange) to $\hat{\gamma}_{n=1}(100\%) = 0.0033$ (red). One can also see again the corresponding down-shift of the pulsation frequency $\omega_{\rm pulse} = 2\omega_{n=1}$ as $\omega_{n=1} \approx \omega_{S0} - \gamma_{n=1}$ obeying Eq.~(\ref{eq:freq_shift_gamma}). After saturation ($\hat{t} > 3000$), Fig.~\ref{fig:09_flux0.5-0.75-1.0_evol}(a) shows that the pulsations are similar in the cases with $\psi_{\rm bnd} = 0.5$ (black dotted) and $0.75$ (orange). In the full-domain simulation, $\psi_{\rm bnd} = 1.0$ (red), the pulsations still have a similar magnitude, but they become more irregular and are elevated by a significant offset. Similarly, the local Fourier spectra in Fig.~\ref{fig:09_flux0.5-0.75-1.0_evol}(c,d,e) show more complexity in the full-domain (e).

Fig.~\ref{fig:09_flux0.5-0.75-1.0_evol}(b) shows the evolution of the mode's peak location $\psi_{\rm peak}$. It starts around $0.03$-$0.04$ in all cases, before increasing to values around $0.04...0.05$ in the saturated regime ($\hat{t} > 3000$). The maximal peak excursion varies somewhat with the domain size, but the significance of these variations is not clear.

The global spectrograms of $\delta B_{\rm tor}$ in Fig.~\ref{fig:09_flux0.5-0.75-1.0_evol}(f,g,h) show similar core-localized quasi-modes on the slow magnetoacoustic continuum plateau in the region $\psi \lesssim 0.1$. In addition, panels (g) and (h) show faint peaks with similar frequency $\omega \approx \omega_{S0}$ near the rational surfaces where $q = m/n = 2,3,...$ has integer values (recall that our simulations are toroidally Fourier-filtered and retain only $|n| = 0,1$, so only integer resonances are to be expected). The details are shown in Figs.~\ref{fig:10_flux0.75cut3_577-evol} and \ref{fig:11_flux1.0_577-evol}.

In the case with $\psi_{\rm bnd} = 0.75$, the $q=2$ surface ($\psi_{q=2} \approx 0.57$) exists in the simulation domain, and Fig.~\ref{fig:10_flux0.75cut3_577-evol} shows that --- in addition to the core-localized slow magnetoacoustic quasi-mode in the $q\sim 1$ region --- a magnetic island with poloidicity $m=2$ appears and disappears periodically at $q=2$. The locations of the X- and O-points alternate between successive pulses, in what appears to be perfect synchronization with the core-localized quasi-mode. Note that the $m/n = 2/1$ tearing mode is stable on its own. This and other useful supplementary information is shown in \ref{apdx:sensitivity_eta-nonunif}.

In the case with $\psi_{\rm bnd} = 1.0$, the simulation domain contains $q=2,3,4,5$ surfaces at $\psi\approx 0.57$, $0.79$, $0.91$ and $0.96$, and Fig.~\ref{fig:11_flux1.0_577-evol} shows that --- in addition to the core-localized slow-magnetoacoustic quasi-mode in the $q\sim 1$ region --- magnetic island with poloidicities $m=2,3,4,5$ appear and disappear periodically at the $q=2,3,4,5$ surfaces. The locations of the X- and O-points seem to at least approximately alternate between successive pulses and they seem to be more or less synchronized with the core-localized slow magnetoacoustic quasi-mode. However, upon close inspection of the Poincar\'{e} plots in panel set (c) one can observe variations in the poloidal phases of the islands. Panel set (e) shows that the pressure perturbations also exhibit variable poloidal phases. Compared to Figs.~\ref{fig:03_flux0.50cut2_577-evol}(a) and \ref{fig:10_flux0.75cut3_577-evol}(a), the time traces of $W_n(t)$ in Fig.~\ref{fig:11_flux1.0_577-evol}(a) are also much more irregular, so the choices we made for the snapshot times (red and blue circles) became somewhat arbitrary.

Although the poorer spatial resolution of the dense flux space in the outer part of the plasma near the $q=4$ and $5$ resonances may play a role, the fact that the mode components in the $q\sim 1.05$ and $q = 2$ regions of the full-domain simulation Fig.~\ref{fig:11_flux1.0_577-evol} undergo noticeable phase changes from one pulse to the next indicates that their dynamic interplay with the $q=3$ resonance (3-body problem) and perhaps even the $q=4,5$ surfaces (many-body problem) are the primary physical cause of the irregularities that we observe in this case.

The coherence time was much longer in the reduced-domain simulations of Figs.~\ref{fig:02_flux0.50cut2_577-evol-exp}--\ref{fig:04_flux0.50_sat-trace} and \ref{fig:10_flux0.75cut3_577-evol}, which may respectively be seen as a one-body problem (isolated core-localized $m/n=1/1$ quasi-mode) and a two-body problem (central $1/1$ quasi-mode and $q=2$ resonance). The fact that phase changes do eventually occur even in those simulations may be linked to the up-down asymmetry of the simulated plasma.

\begin{figure}
	[tbp!]\vspace{-1.6cm}
	\centering
	\includegraphics[width=0.48\textwidth]{\figures/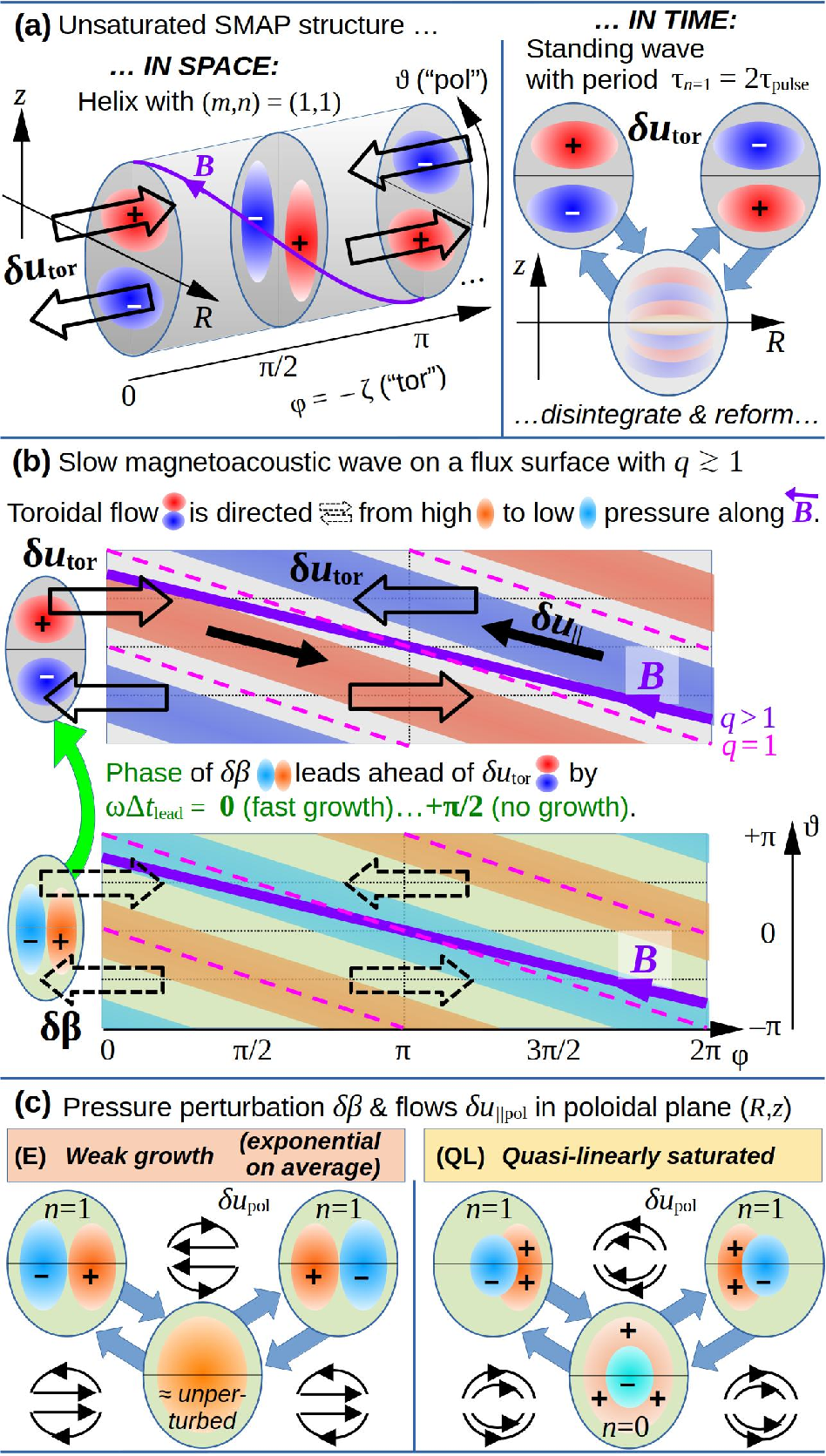}\vspace{-0.1cm}
	\caption{Salient features of slow magnetoacoustic pulsations (SMAPs). (a): Illustration of its spatio-temporal structure in terms of the toroidal flow field $\delta u_{\rm tor}(R,\varphi,z,t)$. The top left panel shows the spatial structure in half of the torus in cylinder coortinates. The quasi-mode is an almost-${\bm B}$-field-aligned helix with dominant poloidal and toroidal mode numbers $(m,n) = (1,1)$. The top right panel illustrates the standing-wave-like pulsations in the poloidal plane $(R,z)$, where the flow direction alternates periodically in the upper and lower half-plane. (b): The oscillation frequency is determined by slow magnetoacoustic waves whose structure on the $(\vartheta,\varphi)$-plane of a magnetic surface with $q \gtrsim 1$ is illustrated here in terms of $\delta u_{\rm tor}(\vartheta,\varphi)$ and $\delta\beta(\vartheta,\varphi)$. Since $q \neq 1$, the $(m,n)=(1,1)$ pressure perturbation $\delta\beta$ (bottom) is not exactly aligned with ${\bm B}$ (violet) and its gradients trigger parallel flows $\delta u_\parallel$ that are directed from high (orange) to low (cyan) $\delta\beta$. Our $\delta u_{\rm tor}$ (top) is the toroidal component of this parallel flow. After eliminating the pressure perturbation, the flow overshoots due to ion inertia and rebuilds the pressure perturbation with opposite sign, and so on. When SMAPs are stable or saturated, $\delta\beta$ oscillations lead ahead of $\delta u_{\rm tor}$ with a phase shift of up to $\omega\Delta t_{\rm lead} \lesssim +\pi/2$. The phase lead tends towards $0$ with increasing exponential growth rate $\gamma$ --- and so does the oscillation frequency, which satisfies $\omega \approx \omega_{\rm S0} - \gamma$ (cf.~Eq.~(\protect\ref{eq:fshift})). (c): During (on-average) exponential growth (E), the pressure peak moves back and forth across the plasma center via the (approximately) unperturbed state as shown here in the $(R,z)$ plane. The phasing of the poloidal flow $\delta u_{\rm pol} \propto \delta\Phi$ relative to $\delta\beta$ varies with the growth rate (Fig.~\protect\ref{fig:07_flux0.50_phase-eta}). Sketched here is the weakly unstable limit, where $\delta u_{\rm pol}$ is ahead of $\delta\beta$, making the mode appear quasi-interchange-like in the poloidal plane, albeit $\delta\beta$ is still mainly controlled by $\delta u_\parallel \approx \delta u_{\rm tor}$ as illustrated in (b). After quasi-linear saturation (QL), the pressure possesses a persistent central depression and annular elevation.}
	\label{fig:12_smap_overview}
\end{figure}

\section{Discussion}
\label{sec:discussion}\vspace{-0.15cm}

\subsection{Summary of observations}
\label{sec:discussion_summary}

\begin{figure}
	[tb!]
	\centering
	\includegraphics[width=0.48\textwidth]{\figures/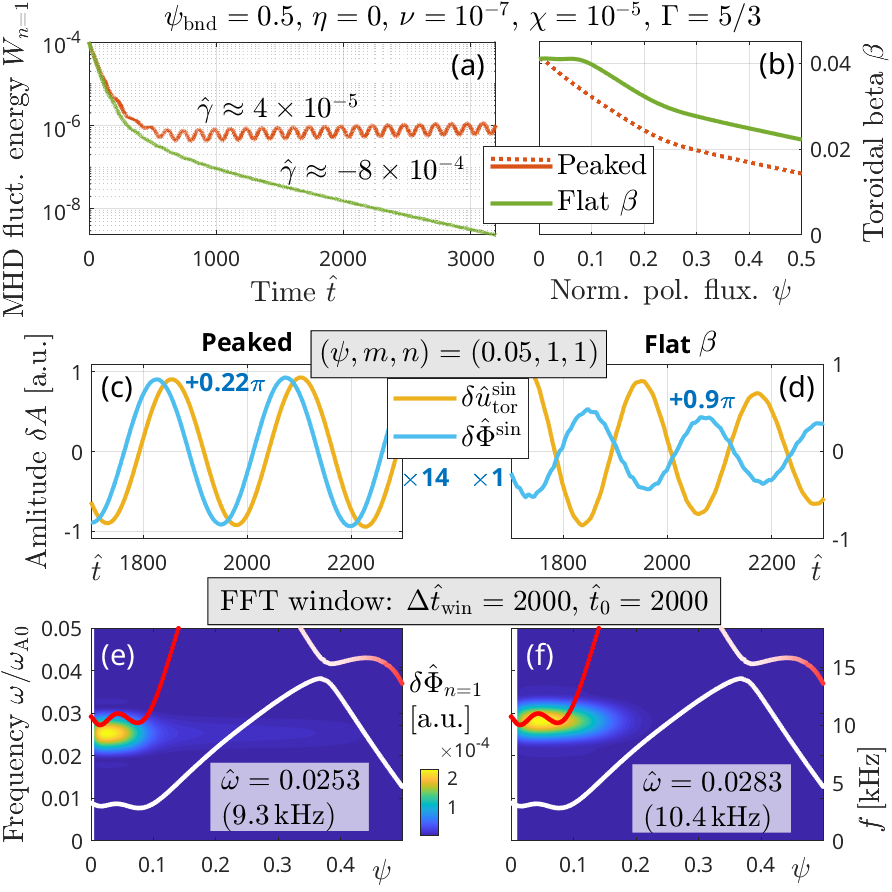}
	\caption{Characterization of low-frequency compressible Alfv\'{e}nic pulsations (LCAPs), found in simulations with zero resistivity ($\eta = 0$), weak viscosity ($\nu = 10^{-7}$), and relatively high thermal diffusivity ($\chi = 10^{-5}$) that eliminates SMAPs (cf.~Fig.~\protect\ref{fig:06_flux0.50cut2_scan-diff}(d)). Panel (a) shows the time traces of the MHD fluctuation energy $W_{n=1}(t)$ obtained, respectively, with the centrally peaked and centrally flat $\beta(\psi)$ profiles that are plotted in panel (b). LCAPs are marginally (un)stable in the peaked $\beta$ case. The pulsation magnitude is much smaller than for SMAPs in Fig.~\protect\ref{fig:06_flux0.50cut2_scan-diff} and becomes barely visible in the flat $\beta$ case. Panels (c) and (d) show the respective waveforms of the $(m,n)=(1,1)$ components of $\delta\Phi$ and $\delta u_{\rm tor}$ measured at $\psi = 0.05$. In panel (c), their amplitude ratio is $|\delta u_{\rm tor}|/|\delta\Phi| \sim 14$, while in panel (d) both quantities are comparable, $|\delta u_{\rm tor}|/|\delta\Phi| \sim 1$, so we plotted the unscaled raw $\delta\Phi$ signal. This shows that the present quasi-mode has a much weaker compressional component than SMAPs in Fig.~\protect\ref{fig:07_flux0.50_phase-eta}(a)--(e). We suspect that the smaller pulsation magnitudes seen in panel (a) is connected with the reduced compressibility (increased Alfv\'{e}nicity) compared to SMAPs. Furthermore, we observe that the two cases exhibit very different phase leads of $\delta\Phi$ relative to $\delta u_{\rm tor}$: in panel (c) there is only a small lead of $\omega\Delta t_{\rm lead} \approx +0.22\pi$, whereas the two waveforms in (d) are nearly out-of-phase with $\omega\Delta t_{\rm lead} \approx +0.9\pi$. Panels (e) and (f) show the respective spectrograms $|\delta\Phi_{n=1}|(\omega,\psi)$. The frequency of the stable LCAPs in (f) match closely the central plateau of the low-frequency Alfv\'{e}nic $(m,n)=(1,1)$ branch. In the marginally (un)stable case (e), the frequency is down-shifted by about $10\%$, which is significantly larger than $\gamma/\omega \approx 3\%$ and thus does not satisfy Eq.~(\protect\ref{eq:fshift}). Further study is required to explain these properties of LCAPs and their practical relevance. Possible coupling to rational surfaces as in Figs.~\protect\ref{fig:10_flux0.75cut3_577-evol} and \protect\ref{fig:11_flux1.0_577-evol}, but at the $\sim 10\,{\rm kHz}$ pace of LCAPs, should be investigated using simulations of the full domain ($\psi_{\rm bnd} = 1$) and including mechanisms for fast, reversible collisionless reconnection (Section~\protect\ref{sec:discussion_nonloc}--\protect\ref{sec:conclusion}).}
	\label{fig:13_flux0.50_Alfvenic-pulse}
\end{figure}

The main observations we made in the simulations described in Section~\ref{sec:results} may be summarized as follows:
\begin{enumerate}
	\item[1.]  {\it SMAP existence and polarization.} In the single-fluid full MHD model, a tokamak plasma with a centrally flat safety factor $q \gtrsim 1$ and sufficiently low effective viscosity and thermal diffusivity ($\nu, \chi < 10^{-5}$) supports pulsations whose pace is determined by the frequency $\omega_{\rm S0}$ of the central plateau of the slow magnetoacoustic continuum's $(m,n)=(1,1)$ branch. The initial perturbation amplitude can be arbitrarily small; there is no amplitude threshold. The associated fluctuations in the pressure and magnetic field satisfy the diamagnetic condition (\ref{eq:diamag_condition}). We call this phenomenon ``slow magnetoacoustic pulsation (SMAP)''.
	
	\item[2.]  {\it SMAP structure.} Salient features of their spatio-temporal structure are summarized schematically in Fig.~\ref{fig:12_smap_overview} and described in the figure's caption. Dominated by poloidal and toroidal Fourier mode numbers $(m,n)=(1,1)$, its spatial form in straight cylinder coordinates as in Fig.~\ref{fig:12_smap_overview}(a) is that of a helical standing wave. As shown in Fig.~\ref{fig:12_smap_overview}(c), the plasma passes through an approximately axisymmetric ($n=0$) configuration between successive $n=1$ pulses of opposite phase, so each full oscillation period of the standing wave comprises two pulses ($\tau_{\rm pulse} = \tau_{n=1}/2$, $\omega_{\rm pulse} = 2\omega_{n=1}$). During the exponentially growing phase (E), the $n=0$ configuration seen between pulses is approximately the unperturbed plasma. After quasi-linear saturation (QL), $n=1$ pulses alternate with an $n=0$ perturbation of similar energy ($W_{n=0} \approx W_{n=1}$).
	
	\item[3.]  {\it Destabilization.} If a pressure gradient exists in the central $q\sim 1$ region, unstable SMAPs are observed in the MHD model when the electric resistivity is sufficiently high (here $\eta > 10^{-6}$). Thus, unlike other resistive instabilities, SMAPs possess a resistivity-dependent instability threshold. Moreover, in the absence of an ambient pressure gradient, resistivity is stabilizing (Fig.~\ref{fig:08_flux0.50_growth-eta}(c)).  During exponential growth, the oscillation frequency $\omega_{n=1}$ is systematically down-shifted from the slow magnetoacoustic continuum plateau $\omega_{\rm S0}$ by an amount similar to the growth rate $\gamma_{n=1}$ (cf.~Eq.~(\ref{eq:freq_shift_gamma})):\vspace{-0.1cm}
	\begin{equation}
		\omega_{n=1} \approx \omega_{\rm S0} - \gamma_{n=1} H(\gamma_{n=1}),
		\label{eq:fshift}\vspace{-0.1cm}
	\end{equation}
	where $H(x)$ is the Heaviside step function, which equals $1$ for $x > 0$, and $0$ otherwise. Apart from this, the SMAP frequency is independent of its amplitude, both during its growth and after saturation.
	
	\item[4.]  {\it Nonlocal effects.} SMAPs can couple to distant rational surfaces all the way to the plasma edge. In our highly resistive simulations, this was found to cause reversible periodic magnetic reconnection with macroscopic magnetic islands waning and waxing, and their X- and O-points appearing at alternating locations during successive cycles.
	
	\item[5.]  {\it Alfv\'{e}nic counterpart.} MHD simulations with negligible resistivity and viscosity (here $\eta = 0$, $\nu = 10^{-7}$) also exhibit pulsations at a higher frequency, slightly below the low-frequency Alfv\'{e}nic continuum plateau's $(m,n)=(1,1)$ branch. We suspect that this is the ``same'' quasi-mode that was studied in Ref.~\cite{Bierwage26a} using antenna drive. In Fig.~\ref{fig:07_flux0.50_phase-eta}(f)--(j) this phenomenon appeared alongside SMAPs. However, unlike the slow magnetoacoustic pulsations studied in this paper, the Alfv\'{e}nic quasi-mode is also seen at higher diffusivity (here $\chi \gtrsim 10^{-5}$), where SMAPs can no longer exist. This is shown in Fig.~\ref{fig:13_flux0.50_Alfvenic-pulse}, whose caption summarizes the salient features. Being located deep in the so-called kinetic thermal ion gap, this low-frequency Alfv\'{e}nic quasi-mode still has a significant compressional component, so we call this phenomenon ``low-frequency compressible Alfv\'{e}nic pulsation (LCAP)''.
\end{enumerate}

In the following subsections, we discuss a few critical questions that arose from the above observations and will make a preliminary attempt to construct a comprehensible physical picture. At present, not all aspects of our observations are fully understood will require further work, even the realm of MHD theory. Meanwhile, the prediction of real-world manifestations of SMAPs and related phenomena like LCAPs, and the assessment of their practical relevance will require future studies with kinetic models. Our current thoughts on this will be outlined in the final Sections~\ref{sec:discussion_nonloc} and \ref{sec:conclusion}.

\subsection{Simultaneous growth \& oscillation}
\label{sec:discussion_qm}

At first glance, the finding of solutions that can grow ($\gamma \neq 0$) and oscillate ($\omega \neq 0$) appears to contradict a fundamental theorem of ideal MHD, whose equations, linearized around a stationary reference state, constitute a Hermitian operator with eigenvalues $\Omega = \omega + i\gamma$ that must be either purely real or purely imaginary \cite{Frieman60}. As we noted near the end of Section~\ref{sec:intro}, we are not aware of cases where the addition of visco-resistive or thermal diffusion violated this ``real-$|\Omega|^2$ theorem''. Two known mechanisms violating this constraint are differential rotation of the background plasma and resonant interactions with particles or other waves (including those coming from an externally applied antenna).

In the case of differentially rotating plasmas, modes with frequencies comparable to or lower than the shearing rate are subject to cyclic phase slippage and realignment. This violates the requirement for a fixed mode structure, giving access to a broader range of solutions.\footnote{A known mathematical description makes use of a dense spectrum of Floquet-type (generalized) eigenmodes with complex eigenvalues and time-dependent wave vectors \protect\cite{Cooper88, Waelbroeck91}, where an unstable modelet is coupled to an infinite number of stable ones \protect\cite{Furukawa05}.\label{fn:floquet}}

Our simulation setup does not involve rotation, but dynamic changes in the mode structure are possible since SMAPs are perturbations on continuous spectra. Continuum modes are known to be time-dependent in the initial-value problem and formally singular in the $t\rightarrow\infty$ eigenvalue problem.\footnote{For instance, shear Alfv\'{e}n modes that are subject to continuum damping \protect\cite{Zonca92,Zonca93,Hu04} exhibit spiky components whose radial width tends asymptotically to zero in the $t\rightarrow\infty$ limit due to radial phase mixing in a spatially nonuniform plasma. This is similar to linear Landau damping in nonuniform velocity distributions. The ideal MHD spikes are resolved by diffusive processes in our model. In kinetic models, this process typically involves mode conversion stages \protect\cite{Chen74,Hasegawa75}.}
Since the continuum damping rate tends to decrease with decreasing magnetic shear $\overline{s} \approx r{\rm d}\ln q/{\rm d}r$, where $r$ is the volume-averaged minor radius, such transients can acquire the form of long-lived quasi-modes in regions where the $q$ profile is nearly flat, so that $\overline{s} \approx 0$. This is precisely the rationale underlying our simulations setup in Fig.~\ref{fig:01_setup}, with centrally flat $q \approx 1.05$. In this way, we stretch the time scale for global initial perturbations to actually become singular so far into the future that they can survive as long-lived global quasi-modes. These may prevail on the millisecond time scale of interest, while pulsating or changing in shape, especially between pulses as seen in Figs.~\ref{fig:02_flux0.50cut2_577-evol-exp}(e,g) and \ref{fig:04_flux0.50_sat-trace}(h).

However, one may argue that the radial phase mixing of continuum modes does not truly violate the real-$|\Omega|^2$ theorem because each singular wave on its own is purely oscillatory ($\gamma = 0$) and it is merely the envelope that decays, while total wave energy is conserved. Thus, the flexible form of the fluctuation envelope on a continuum plateau is not a satisfactory explanation on its own.

Our current understanding of how SMAPs violate the real-$|\Omega|^2$ theorem is that they consist of two modes --- one transverse ($\perp$) and one parallel ($\,\parallel\,$) to the magnetic field ${\bm B}$ --- whose synergy gives rise to an oscillatory instability by providing a mechanism to tap the energy stored in an ambient pressure gradient:
\begin{enumerate}
	\item[($\perp$)] resistive quasi-interchange, and
	\item[($\,\parallel\,$)] slow magnetoacoustic waves with high radial coherence facilitated by a $q \sim 1$ continuum plateau.
\end{enumerate}

\noindent Their interaction is reflected in the observed frequency shift (\ref{eq:fshift}) and phase relations between three different field components ($\delta u_{\rm rad} \propto \delta\Phi$, $\delta u_{\rm tor}$, $\delta\beta$) in Figs.~\ref{fig:02_flux0.50cut2_577-evol-exp}(a) and \ref{fig:07_flux0.50_phase-eta}, which both vary with the growth rate $\gamma$ that, in turn, depends on the resistivity $\eta$ (and other parameters).

This interaction may be seen as a form of ``resonance'', which is one of the known pathways to violate the real-$|\Omega|^2$ theorem. The complete physical picture as we see it now is described in the following Section~\ref{sec:discussion_n01}. 

\begin{figure*}
	[tb!]\vspace{-2.7cm}
	\centering
	\includegraphics[width=0.96\textwidth]{\figures/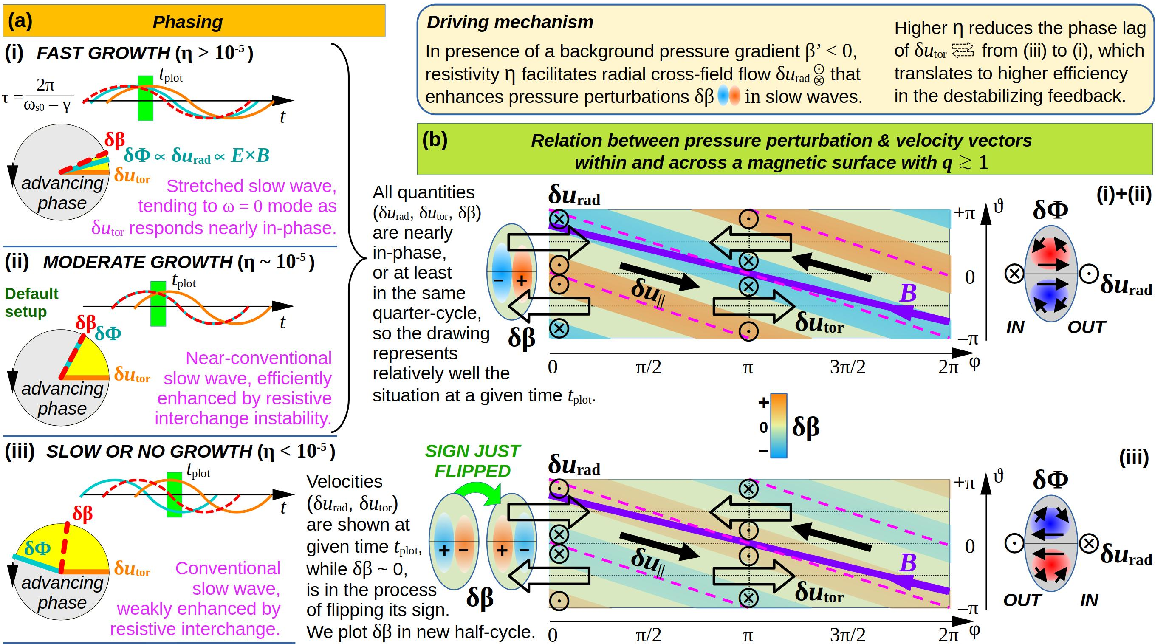}\vspace{-0.15cm}
	\caption{Mechanism of SMAP destabilization. Column (a) summarizes schematically the relative phasing of oscillations in three relevant fields --- $\delta u_{\rm tor}$ (orange), $\delta\Phi \propto \delta u_{\rm rad} \propto {\bm E}\times{\bm B}$ (cyan), and $\delta\beta$ (red dashed) --- as observed in Figs.~\protect\ref{fig:02_flux0.50cut2_577-evol-exp}(a) and \protect\ref{fig:07_flux0.50_phase-eta}. As before, we use the toroidal velocity $\delta u_{\rm tor}$ as a reference since its waveform is more robust than that of the radial component $\delta\Phi \propto \delta u_{\rm rad}$ (as we saw in Fig.~\protect\ref{fig:04_flux0.50_sat-trace}). Here, $\delta u_{\rm tor}$ is primarily the toroidal component of the parallel velocity $\delta u_\parallel$. Together with the (normalized) pressure perturbation $\delta\beta$, $\delta u_{\rm tor}$ forms the parallel slow magnetoacoustic wave component of SMAPs, while $\delta u_{\rm rad}$ represents the transverse Alfv\'{e}nic (electromagnetic) component. We choose to distinguish between cases with (i) fast, (ii) moderate, and (iii) slow or no growth, whose main characteristics are described in the respective magenta text segments of column (a). The fast and slowly growing cases (i) and (iii) correspond, respectively, to the results obtained with $\eta = 5\times 10^{-5}$ and $10^{-6}$ in Figs.~\protect\ref{fig:06_flux0.50cut2_scan-diff} and \protect\ref{fig:07_flux0.50_phase-eta}. The moderate case (ii) is our default scenario, details of which where presented in Figs.~\protect\ref{fig:02_flux0.50cut2_577-evol-exp}--\protect\ref{fig:04_flux0.50_sat-trace}. Column (b) illustrates the corresponding spatial structure of the three fields within and across a magnetic surface with $q \gtrsim 1$, using the same drawing conventions and coordinates as in Fig.~\protect\ref{fig:12_smap_overview}. Note that column (b) is able to reflect only the relative sign (in- or out-of-phase) of the three fields at a chosen time, not their exact phasing. The situation for fast and moderate growth (i) and (ii) is summarized in a single plot because all fields have similar phases (i) or are at least in the same quarter-cycle (ii). The approximate snapshot time $t_{\rm plot}$ is indicated by a green rectangle in each row of column (a). In the slowly growing or marginal case (iii), the velocities $\delta\Phi\propto\delta u_{\rm rad}$ and $\delta u_{\rm tor}$ are nearly out-of-phase, so one key difference between the upper and lower plots in column (b) is the opposite orientation of the radial flow $\delta u_{\rm rad}$ indicated by $\odot$ and $\otimes$ symbols. The second difference is that the phase of $\delta\beta$ lies half-way between those velocities, so it passes through zero and changes its orientation around our chosen plotting time $t_{\rm plot}$. Here, we assume that the sign has just flipped, so we plot $\delta\beta(\vartheta,\varphi)$ for the new half-cycle, where $\delta\beta$ has the same sign as $\delta u_{\rm rad}$ (while lagging nearly $\pi/2$ behind), and where $\delta u_{\rm tor}$ has entered the overshoot phase (enhancing $\delta\beta$ rather than reducing it). The key observation is that the phase relation of $\delta\beta$ (orange/cyan) and $\delta u_{\rm rad}$ ($\odot$/$\otimes$) is such that, on average, the radial cross-field flow acts to enhance the pressure perturbations as summarized in the text box at the top.}
	\label{fig:14_smap_instability}\vspace{-0.35cm}
\end{figure*}

\subsection{Physical picture}
\label{sec:discussion_n01}

At present, our interpretation of the results presented in this paper leads us to conclude that the primary mechanism facilitating exponential growth of SMAPs is\vspace{-0.05cm}
\begin{itemize}
	\item  pressure-gradient-driven resistive ``interchange'' or (since we have magnetic curvature) ``ballooning''.
\end{itemize}

\noindent Here, it is the nonzero electric resistivity $\eta$ that allows the mode to tap an existing background pressure gradient via diffusive transport across magnetic surfaces. The utter lack of power-law behavior in Fig.~\ref{fig:08_flux0.50_growth-eta} --- especially a strong deviation from the $\eta^{1/3}$ scaling predicted for conventional resistive interchange instabilities \cite{Furth63, Carreras87} --- is not considered to be problematic for our interpretation. We think that this is due to the same reason as the violation of the real-$|\Omega|^2$ theorem discussed in Section~\ref{sec:discussion_qm} above: SMAPs are compound modes integrating non-ideal transverse and ideal parallel dynamics, where the $\eta$-dependence enters in intricate ways that affect the mode structure and polarization (to such a degree that SMAPs even have an $\eta$-dependent instability threshold).

Overall, exponentially growing SMAPs resemble alternating quasi-interchange dynamics, where the attribute ``quasi-'' highlights the fact that there are not only dynamics transverse to ${\bm B}$ but also parallel; namely, the slow magnetoacoustic waves. The latter are usually dominant in SMAPs. The relative strength of the transverse interchange component and the parallel acoustic component is reflected in several measurable quantities. One is the frequency shift $\omega = \omega_{\rm S0} - \gamma$ in Eq.~(\ref{eq:fshift}). Others are ratios like $|\delta u_{\rm rad}|/|\delta u_{\rm tor}| \propto |\delta\Phi|/|\delta\beta|$ that represent inverse Alfv\'{e}nicity,\footnote{A more refined measure is the Alfv\'{e}nicity parameter $\A \equiv G_1^2/(G_1^2 + G_2^2) \in [0,1]$ in Eq.~(2.9) or Ref.~\protect\cite{Falessi20}, where $G_1 \propto \delta\Phi$ and $G_2 \propto \delta\beta$ \protect\cite{Falessi19b}. This parameter $\A$ determines the coloring of the MHD continua shown in our spectrograms like Fig.~\protect\ref{fig:12_smap_overview}(e,f), where red is the Alfv\'{e}nic ($\A \rightarrow 1$) and white the acoustic limit ($\A \rightarrow 0$).}
and which can be inferred from Fig.~\ref{fig:04_flux0.50_sat-trace} as wells as from the scaling factors printed blue in Figs.~\ref{fig:02_flux0.50cut2_577-evol-exp}(a) and \ref{fig:07_flux0.50_phase-eta}.
With increasing growth rate, SMAPs acquire a stronger (though still very weak) Alfv\'{e}nic/electromagnetic component and their frequency decreases from $\omega_{\rm S0}$ of acoustic oscillations towards zero (limit of purely growing modes).

Figs.~\ref{fig:02_flux0.50cut2_577-evol-exp}(a) and \ref{fig:07_flux0.50_phase-eta} showed that not only the growth rate $\gamma$ but also the phasing between $\delta u_{\rm rad} \propto \delta\Phi$, $\delta u_{\rm tor}$, and $\delta\beta$ varied with the resistivity $\eta$. As is known from other instabilities, like resistive drift waves, such phase relations determine how a mode extracts and converts free energy from ambient gradients. Our phase analysis is illustrated schematically in Fig.~\ref{fig:14_smap_instability}, where the text box at the top summarizes our conclusion concerning the mechanism of SMAP destabilization. 

This insight is embedded in our current physical picture of SMAPs, which may be summarized as follows. The first bullet is partly speculatory, while the rest is largely evidence-based (though but readers are encouraged to view all our interpretations critically):
\begin{itemize}
	\item  Slow waves on neighboring magnetic surfaces are more or less synchronized since the continuum plateau lacks radial dispersion (${\rm d}\omega_{\rm S}/{\rm d}r \approx 0$ for $q \sim 1$). This radial synchronization facilitates coherent global convective cells in the poloidal plane.
	
	\item  Co- and counter-propagating ($\omega_{\rm S} \lessgtr 0$) slow waves produce a standing $(m,n)=(1,1)$ wave, so that the associated global convective cells stay in place while their circulation direction alternates in time.
	
	\item  In the presence of a pressure gradient in the $q\sim 1$ region, these alternating convective cells move the central pressure peak back and forth across the plasma center. The analysis in Fig.~\ref{fig:14_smap_instability} confirmed that the phases of various field components are always such that, on average, a negative ambient pressure gradient $\beta' \equiv {\rm d}\beta/{\rm d}r$ enhances the pressure perturbation $\delta\beta$ in the slow waves, thus, causing them to grow. The quantitative phase difference, between $\delta\Phi\propto\delta u_{\rm rad}$, $\delta u_{\rm tor}$ and $\delta\beta$ determines the efficiency of the positive feedback and is controlled here by the resistivity $\eta$ since it controls the possible rate of cross-field transport and thus energy conversion. Since resistive dissipation is irreversible, the energy released from the flattening of the ambient pressure gradient accumulates in the slow waves, allowing them to grow until the drive has been depleted to such a degree that it can no longer compete with damping mechanisms.\footnote{See~\ref{apdx:sensitivity_eta-heating} for a possible additional twist with respect to the role of visco-resistive heating in Eq.~(\protect\ref{eq:model_cc_pre}) of our model.}
	
	\item  The similarity of SMAP waveforms observed during the exponentially growing and saturated regimes, and the lack of higher harmonics $|n|\geq 2$ tells us that the saturation mechanism is a quasi-linear one; here, the depletion of the ambient pressure gradient. This is further corroborated by Fig.~\ref{fig:15_flux0.50_filt-n0}, which shows that SMAPs fail to saturate when the $n=0$ Fourier component filtered out.
	
	\begin{figure}
		[tbp!]\vspace{-0.7cm}
		\centering
		\includegraphics[width=0.48\textwidth]{\figures/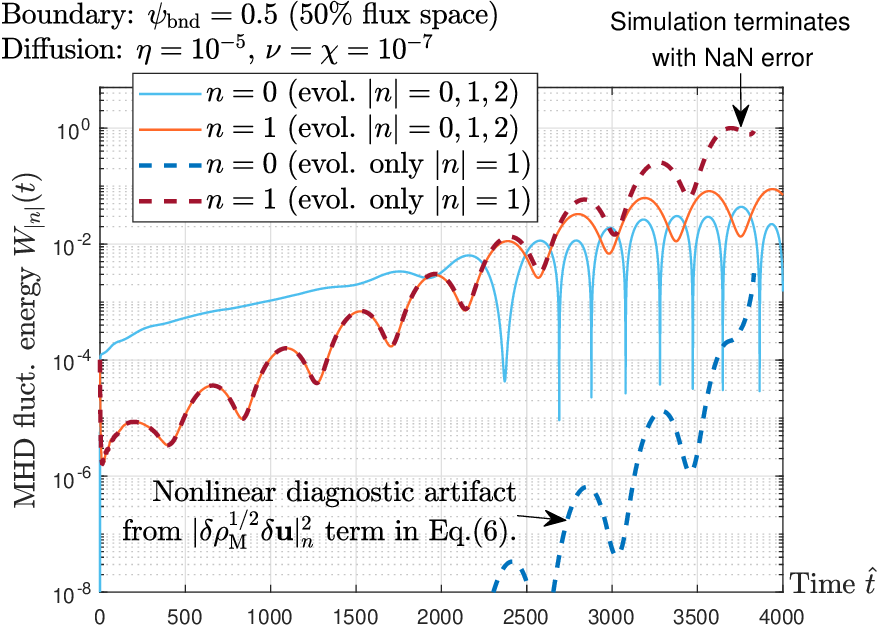}
		\caption{Demonstration that the $n=0$ component in Fig.~\protect\ref{fig:03_flux0.50cut2_577-evol} is required for quasi-linear saturation around $\hat{t} \approx 3000$ and beyond, but plays no significant role during the exponential growth ($\hat{t} < 2500$) of SMAPs. The solid lines are from Fig.~\ref{fig:03_flux0.50cut2_577-evol}, where we evolved $|n| = 0,1,2$. The dashed lines were obtained when all toroidal Fourier components except $|n|=1$ were filtered out at each time step.}
		\label{fig:15_flux0.50_filt-n0}%
	\end{figure}
	
	\item  In our default simulation setup with weak viscous and diffusive damping ($\nu = \chi = 10^{-7}$), SMAPs persist as nearly harmonic oscillations after saturation. Saturated SMAPs seem to have lost much of their radially convective component and are henceforth dominated by standing slow magnetoacoustic waves with $\omega \approx k_\parallel v_{\rm S0}$. Their co- and counter-propagating components cyclically accumulate and disperse a helical $(m,n) = (1,1)$ pressure perturbation on a flux surface. Between successive $|n|=1$ pulses, part of the helical structure is thus temporarily converted into a $(0,0)$ perturbation. This interplay between $n=0$ and $|n|=1$ via parallel streaming $\delta u_\parallel \approx \delta u_{\rm tor}$ explains the out-of-phase pulsations in the volume-integrated energies $W_{n=0}$ and $W_{|n|=1}$ that are characteristic for SMAPs.\footnote{While seen most clearly in the saturated regime in Fig.~\ref{fig:03_flux0.50cut2_577-evol}(a), the $n=0$ pulsations can also be detected during the exponential growth phase if on subtracts the ``heat'' that was generated from the dissipation of early transients and dominates $W_{n=0}$ at low amplitudes.}
\end{itemize}

In the following, we discuss some of the critical ingredients required for SMAP instability and existence. In doing so, we incorporate brief cursory surveys of the existing literature, with no claims of completeness.

\subsection{Role of Alv\'{e}nization \& compressibility}
\label{sec:discussion_smai}

Much of the literature of magnetoacoustic instabilities seems to deal with high-frequency phenomena also known as magnetoacoustic cyclotron instabilities (MCI), which is resonantly driven by the bump-on-tail mechanism and which is thought to be a source of ion cyclotron emission (ICE) observed in many experiments. We found a few works dealing explicitly with {\it slow} magnetoacoustic waves, whose instability has been linked to some kind of ``coupling'' with Alfv\'{e}n waves:
\begin{itemize}
	\item McKenzie \& Webb 1984 \cite{McKenzie84} find that ``{\it backward propagating slow magnetoacoustic waves can be driven unstable by the pressure of the self-excited Alfv\'{e}n waves.}''\vspace{-0.05cm}
	
	\item Rubtsov, Mager \& Klimushkin 2018 \cite{Rubtsov18} show that ``{\it the instability develops on the slow magnetoacoustic oscillation branch, but the instability threshold is determined by the coupling with the Alfv\'{e}n mode.}''\vspace{-0.03cm}
\end{itemize}

\noindent In Fig.~\ref{fig:07_flux0.50_phase-eta}(a)--(e) we saw that the Alfv\'{e}nicity of SMAPs is very weak, but the ratio $\propto|\delta\Phi|/|\delta u_{\rm tor}|$ does increase with increasing SMAP growth rate $\gamma_{n=1}$. It is currently unclear to us whether this change in our quasi-mode's polarization is usefully described as some kind of ``coupling'' to Alfv\'{e}n waves as done in the above references. The appropriate physical picture may depend on the particular system studied. The above references deal with space plasmas, which have very different geometries and plasma parameters compared to our magnetically confined fusion (MCF) plasmas.

In tokamaks and stellarators, one speaks of coupling between the two branches when their continuous spectra develop Alfv\'{e}n-acoustic gaps \cite{Cheng86} like the one in Fig.~\ref{fig:01_setup}(d). Discrete modes can be found in or near those gaps \cite{Gorelenkov07a, Bierwage15b} but we are not aware or reports where such modes became unstable in the absence of kinetic effects (in particular, wave-particle resonances).

Thus, in the case of MCF plasmas --- where SMAPs are so far known to exist only in our simulations and await experimental confirmation --- we propose to speak of a ``(weak) Alfv\'{e}nization'' of the slow magnetoacoustic branch, which happens independently of the coexisting Alfv\'{e}nic branch. These may be mere semantics, but they matter for clarity because the concept of Alfv\'{e}n-acoustic coupling is already linked to other MCF plasma phenomena (which may play no role in space plasmas).

The coexistence of acoustic and Alfv\'{e}nic pulsations was demonstrated in Fig.~\ref{fig:07_flux0.50_phase-eta}(f)--(j). Both quasi-modes seem to have mixed polarization, and Fig.~\ref{fig:13_flux0.50_Alfvenic-pulse} shows that the pulsations weaken with increasing Alfv\'{e}nicity, suggesting that compressibility is important for pulsations.

\subsection{Feasiblity of thermal misbalance}
\label{sec:discussion_exist}

At the time of this writing, AI-generated summaries of the literature tell us that slow magnetoacoustic instabilities are caused by ``thermal misbalance'' and that ``field-aligned thermal conduction ($\propto \chi_\parallel^{-1}$) acts to stabilize both slow and entropy modes.'' The stated role of thermal diffusivity $\chi$ is consistent with the results of our simulations. Meanwhile, we have not investigated whether entropy waves play a distinct role here. On the one hand, it is possible that the observed facilitating role of low thermal diffusivity $\chi$ for the existence of SMAPs in our simulations may be interpreted as evidence for some kind of ``coupling'' to entropy waves. For instance,\vspace{-0.05cm}
\begin{itemize}
	\item Kolotov, Nakariakov \& Fihosy 2023 \cite{Kolotkov23} write in the context of solar coronal plasma that ``{\it The effect of the back-reaction of the wave-induced thermal misbalance on coronal MHD waves has recently become subject to intensive studies. In particular, the misbalance was found to cause coupling of slow magnetoacoustic and entropy waves.}''\vspace{-0.05cm}
\end{itemize}

\noindent On the other hand, our current understanding is that the concept of entropy waves refers to the presence of thermal nonuniformities (``hot and cold spots'') that are primarily convected with the fluid rather than propagating as independent waves. This leaves us with the impression that entropy waves and thermal misbalance may be two names for the same phenomenon, at least within the scope of our Eqs.~(\ref{eq:model_cc}) and (\ref{eq:model_cc_alg}). If this is the case, the concept of entropy waves may not need to be invoked here. Instead, we may assert that SMAPs can arise when slow magnetoacoustic waves $\omega_{\rm S}$ are sustained by mechanisms that sustain thermal nonuniformities (misbalance) on a similar or longer time scale $\tau \gtrsim 2\pi/\omega_{\rm S}$.

The next question is the feasibility of thermal misbalance in the core of a fusion-relevant tokamak plasma. In our simulations, low values of the effective viscosity $\nu$ and thermal diffusivity $\chi$ of the MHD fluid ensure, respectively, that ${\bm E}\times{\bm B}$ flows are only weakly damped and thermal misbalance can be maintained at long wavelengths and low frequencies. We found that these factors are necessary prerequisites for the existence of SMAPs.

In a weakly collisional plasma, the effective viscosity $\nu$ and perpendicular diffusivity $\chi_\perp$ of the ions is determined primarily by ``neoclassical'' effects (mirror force, finite gyroradii and magnetic drifts) and fluctuation-induced ``anomalous'' transport. Their influence is difficult to estimate accurately, especially near the magnetic axis, but we shall assume that our requirement $\nu,\chi_\perp \lesssim 10^{-6}$ is at least a plausible scenario for the core of a tokamak. The primary source of doubt for the existence of SMAPs is the parallel diffusivity $\chi_\parallel$ that controls the sustainability of thermal misbalance/nonuniformity along the magnetic field ${\bm B}$, so we devote the following paragraphs to a preliminary assessment of its effect in the core of a tokamak plasma with $q \sim 1$.

According to Fig.~\ref{fig:05_flux0.50cut2_scan-G}, SMAPs consist of slow magnetosonic waves, so we invoke the dispersion relation\vspace{-0.2cm}
\begin{equation}
	\omega_{\rm S} = k_\parallel c_{\rm S},\vspace{-0.2cm}
\end{equation}

\noindent where the speed of sound $c_{\rm S}$ of a quasi-neutral deuterium plasma with number densities $Z_{\rm D}N_{\rm D} + Z_{\rm e} N_{\rm e} \approx 0$ (with $-Z_{\rm e} = Z_{\rm D} = 1$), electron and ion temperatures $T_{\rm e}$ and $T_{\rm D}$, and specific heat ratio $\Gamma$,\footnote{Electrons are often assumed to be isothermal ($\Gamma_{\rm e} \approx 1$) and ions adiabatic ($\Gamma_{\rm i}^{\rm 1D} = 3$, $\Gamma_{\rm i}^{\rm 3D} = 5/3$. In single-fluid MHD, there is no distinction between the species and the conventional (albeit sometimes disputed) choice is $\Gamma = 5/3$. For our rough estimates, we adhere to this convention while considering the possibility of $T_{\rm e} \neq T_{\rm D}$.}
can be estimated as\vspace{-0.05cm}
\begin{equation}
	c_{\rm S}^2 = \frac{\Gamma_{\rm e} T_{\rm e} + \Gamma_{\rm D} T_{\rm D}}{M_{\rm D}} = \frac{\Gamma_{\rm e} P_{\rm e} + \Gamma_{\rm D} P_{\rm D}}{N_{\rm D} M_{\rm D}} \sim \frac{\Gamma \beta}{2} v_{\rm A}^2
	\label{eq:cs}\vspace{-0.05cm}
\end{equation}

\noindent Balancing ($\leftrightarrow$) the time derivative on the left-hand side of Eqs.~(\ref{eq:model_cc_den}) and (\ref{eq:model_cc_pre}) with the diffusion term on the right-hand side in the direction parallel to ${\bm B}$ gives the condition\vspace{-0.35cm}
\begin{equation}
	\omega_{\rm S} \;\; \leftrightarrow \;\; \chi_{\rm th,\parallel} k_\parallel^2 = \chi_{\rm th,\parallel} \frac{\omega_{\rm S}^2}{c_{\rm S}^2} = \chi_{\rm th,\parallel} \frac{2\omega_{\rm S}^2}{\Gamma \beta v_{\rm A}^2}.
	\label{eq:chipar_balance}\vspace{-0.05cm}
\end{equation}

\noindent Applying our normalizations and substituting the on-axis values $\beta_0 = 4\%$ and $\hat{\omega}_{\rm S0} = 0.0083$ for our KSTAR-based example, and letting $\Gamma = 5/3$, we obtain\vspace{-0.05cm}
\begin{equation}
	\chi_\parallel \;\; \leftrightarrow \;\; \frac{\Gamma\beta_0}{2\hat{\omega}_{\rm S0}} = \frac{5\times 0.04}{6\times 0.0083} \approx 4.
	\label{eq:chipar_value}\vspace{-0.05cm}
\end{equation}

\noindent This is the value of $\chi_\parallel$ at which parallel thermal diffusion would be able to compete with the mode's oscillations.

However, Fig.~\ref{fig:06_flux0.50cut2_scan-diff}(e) shows that the pulsations disappear already for $\chi > 5\times 10^{-6}$, which is 6 orders of magnitude smaller than the level at which Eq.~(\ref{eq:chipar_value}) predicts parallel diffusion to become dominant. This suggests that the disappearance of pulsations in Fig.~\ref{fig:06_flux0.50cut2_scan-diff}(e) for $\chi > 5\times 10^{-6}$ is caused by perpendicular diffusion of the $m/n=1/1$ component near the magnetic axis, where its poloidal wavenumber $k_{\rm pol} = m/r$ becomes large. Indeed. the perpendicular version of condition (\ref{eq:chipar_balance}) is\vspace{-0.05cm}
\begin{equation}
	\hat{\omega}_{\rm S0} = 0.0083 \;\; \leftrightarrow \;\; \chi_\perp \frac{R_0^2}{r^2} \approx 10^{-5} \left(\frac{1.81\,{\rm m}}{0.6\,{\rm m}}\right)^2 \frac{a^2}{r^2},\vspace{-0.05cm}
\end{equation}

\noindent which is satisfied for minor radii around $\hat{r} \equiv r/a \approx 0.1$. This implies that for $\chi_\perp \gtrsim 10^{-5}$, transverse diffusion competes with the mode's oscillation across a significant portion of the central $q\ \sim 1$ plateau in Fig.~\ref{fig:01_setup}, which may explain the results in Fig.~\ref{fig:06_flux0.50cut2_scan-diff}(e).

As discussed at the beginning of this subsection, we shall be content with assuming that $\chi_\perp \lesssim 10^{-6}$ can be a realistic representation of a tokamak plasma, but it still remains to ascertain that $\chi_\parallel \ll 4$ can be realized in practice, so that the situation in Eq.~(\ref{eq:chipar_value}) is avoided and SMAPs can exist. For the following preliminary estimates, we shall leave behind the single fluid MHD model and consider the two-fluid or kinetic picture.

If $T_{\rm e} \gg T_{\rm D}$, we have $P \approx P_{\rm e}$ and the high parallel heat conductivity of electrons (anisotropy ratios commonly cited in the literature are around $\chi_\parallel/\chi_\perp \sim 10^{10}...10^{12}$) makes parallel temperature fluctuations practically negligible. Consequently, pressure fluctuations $\delta P \approx T_{\rm e}\delta N_{\rm D}$ are mainly caused by ion density fluctuations. Moreover, the sound speed $c_{\rm S}$ in Eq.~(\ref{eq:cs}) becomes insensitive to $T_{\rm D}$ when $T_{\rm e} \gg T_{\rm D}$, so that the ion thermal velocity $v_{\rm th,0} = \sqrt{2T_{\rm D0}/M_{\rm D}}$ can become significantly smaller than $c_{\rm S}$. If this is the case, the effect of ion Landau damping on slow magnetoacoustic waves becomes weak. Thus, it is possible that SMAPs may occur in tokamak plasmas with dominant electron heating.

In the KSTAR deuterium plasma on which our simulations are based, $T_{\rm e0} \approx 4\,{\rm keV}$ and $T_{\rm D0} \approx 3\,{\rm keV}$ were comparable. In this case, the following rough estimates remain inconclusive about the existence of SMAPs. A thermal deuteron with an energy of $3\,{\rm keV}$ has a speed of $v_{\rm th,0} \approx 536\,{\rm km/s}$. If $|v_\parallel| = v_{\rm th}$, this particle will perform a toroidal turn around our KSTAR plasma in approximately $0.02\,{\rm ms}$, or at rate of $50\,{\rm kHz}$. Since our waves oscillate at $\omega_{\rm S0} \approx 2\pi\times 3\,{\rm kHz}$, such a deuteron will encircle the plasma about $16$ times during one slow magnetoacoustic oscillation period. Most particles in a near-Maxwellian distribution, especially mirror-trapped ones, have $|v_\parallel| < v_{\rm th}$, so they will make fewer turns during a wave period. Meanwhile, with $q_0 \approx 1.05$, about $N_{2\pi} \sim 1/|1-q_0| \approx 20$ toroidal turns are required in order to poloidally phase-average the $(m,n) = (1,1)$ quasi-mode. The value of $N_{2\pi} \sim 20$ lies in the same ballpark as the above $v_{\rm th,0}/(R\omega_{\rm S0}) \sim 16$. Clearly, the mode is likely to experience significant damping, but whether this is sufficient to entirely eliminate SMAPs is not immediately evident and remains to be tested numerically, using kinetic models of the plasma.
\subsection{Nonlocal \& multi-scale effects}
\label{sec:discussion_nonloc}

The present series of papers is dedicated to the study of ``low-frequency core-edge coupling in tokamaks'', and the present study of core-localized SMAPs was included in this series because we found that SMAPs can induce magnetic reconnection at distant radii, all the way to the plasma edge. The relevant observations were summarized in Figs.~\ref{fig:10_flux0.75cut3_577-evol} and \ref{fig:11_flux1.0_577-evol}.

This is somewhat reminiscent of so-called resonant tearing modes (rTM), that were studied for instance by Cai \& Ding \cite{CaiH22} and Yu {\it et al}.~\cite{YuLM24} in the presence of fast ions. rTMs can arise during the interval after a sawtooth crash and before the onset of fishbone oscillations.

In our case, the drive comes from a pressure gradient in the thermal bulk plasma that feeds weakly Alfv\'{e}nized slow magnetoacoustic waves, which then couple through the compressible toroidal plasma to distant integer rational surfaces.

The reverse process of edge-to-core coupling was studied using antenna drive in the previous paper of this series \cite{Bierwage26a} using nearly the same plasma conditions. For the Alv\'{e}nic branch ($\sim 10\,{\rm kHz}$) studied in Ref.~\cite{Bierwage26a}, we demonstrated that volumetric focusing of wave energy plays an important role for driving efficiency. In the present case of core-to-edge coupling, radial energy transfer is subject to the reverse process: volumetric dilution. It is interesting that, at least at the low acoustic frequencies ($\sim 3\,{\rm kHz}$) we are dealing with here, the coupling is sufficiently efficient to drive reconnection.

While the ``reversibility'' of magnetic reconnection involving macroscopic magnetic islands as seen in Figs.~\ref{fig:10_flux0.75cut3_577-evol} and \ref{fig:11_flux1.0_577-evol} is facilitated here by a high resistivity and correspondingly high loop voltage, kinetic plasma models have other channels to redistribute magnetic flux. One collisionless mechanism that is considered to be relevant in tokamaks is electron inertia, whose effect was demonstrated in extended MHD \cite{Ottaviani95, Bierwage07} and gyrokinetic simulations \cite{Ishizawa13}.

In contrast to the resistivity-mediated case, collisionless reconnection is truly reversible\footnote{Electron-inertia-mediated ``reconnection'' changes magnetic topology (magnetic flux surfaces) while preserving the topology of a generalized flux function, also known as canonical helicity and related to the canonical momentum of electrons \protect\cite{Ottaviani93, Schep94, Ottaviani95, Cafaro98, Yoon18}. Similarly, the two-fluid model possesses a generalized flux with conserved topology when electron pressure effects are added \cite{Cafaro98}. Consequently, these are conservative (Hamiltonian) and thus reversible processes. Irreversible topology change requires some kind of (anomalous) diffusion \protect\cite{Drake91}.}
and ``fast'',\footnote{With respect to fast reconnection, Wesson \protect\cite{Wesson90} pointed out that ``{\it the acceleration of the electrons constitutes a larger impedance than the collisional resistivity. This broadens the current layer and allows a greater flow of plasma through the layer}''.}
so it may be driven also by modes with higher frequencies than those studied here. This may be of interest for the interpretation of tearing parity signatures seen in ``multi-scale modes'' on DIII-D that oscillate in the $\lesssim 20\,{\rm kHz}$ range in the plasma frame as reported by Du {\it et al}.~\cite{Du21}. These frequencies may be too high for SMAPs, but their Alfv\'{e}nic counterpart LCAP in Fig.~\ref{fig:13_flux0.50_Alfvenic-pulse} could play a role.

\section{Conclusion}
\label{sec:conclusion}

In this paper, we reported the finding of spontaneously excited magnetohydrodynamic (MHD) pulsations in the central core of a tokamak plasma with centrally flat safety factor slightly above unity ($q \gtrsim 1$) and relatively high normalized pressure of $\beta \sim 4\%$. The simulation scenario resembles the size and geometry of a KSTAR tokamak plasma in the so-called hybrid scenario. In this configuration, the low-frequency MHD continua with poloidal and toroidal mode numbers $(m,n) = (1,1)$ have flat plateaus that cover about $20\%$ of the torus' minor radius $r$. The said pulsations were found on the central plateaus of both the acoustic branch ($\omega_{\rm S0} \approx 3\,{\rm kHz}$) and the low-frequency Alfv\'{e}nic branch ($\omega_{\rm A0} \approx 10\,{\rm kHz}$), albeit under different conditions (to be described shortly).

In the case of the acoustic branch, which was the main focus of the present paper, we speak of ``slow magnetoacoustic pulsations (SMAPs)''. Based on detailed numerical analyses, sensitivity studies, and parameter scans, we constructed a physical picture of SMAPs that was described in Section~\ref{sec:discussion_n01} and illustrated in Figs.~\ref{fig:12_smap_overview} and \ref{fig:14_smap_instability}. In a nutshell, we interpret SMAPs as arising from the interplay between radially synchronized slow magnetoacoustic standing waves that act as a pacemaker, and resistive quasi-interchange (or ``ballooning'' since we have magnetic curvature) that acts as an energy converter. Sourced by an ambient pressure gradient $\beta' \equiv {\rm d}\beta/{\rm d}r$, and with resistivity $\eta$ allowing for cross-field radial transport, SMAPs in our simulations grew exponentially on average and saturated quasi-linearly when the energy originally stored in the ambient pressure gradient was exhausted. The pulsations continued after saturation in the form of nearly undamped magnetoacoustic waves when viscous damping and thermal diffusion are both sufficiently weak.

While crucial for their instability, neither the pressure gradient $\beta'$ nor resistivity $\eta$ are necessary for the existence of (stable) SMAPs as such. Nevertheless, when nonzero, these parameters can noticeably alter the oscillation frequency via the growth rate $\gamma_{n=1}$ as shown in Eq.~(\ref{eq:fshift}): $\omega_{n=1} \approx \omega_{\rm S0} - \gamma_{n=1}$. With increasing drive, the mode tends towards a conventional zero-frequency purely growing MHD instability (here resistive quasi-interchange/ballooning). When drive is weak, the mode properties are dominated by the acoustic waves on their central continuum plateau where $q\sim 1$.

While resistivity $\eta$ is destabilizing in the presence of a central pressure gradient, increasing $\eta$ was found to be stabilizing in simulations with a centrally flat pressure profile. This corroborates the fact that resistivity does not act directly on the slow waves (traveling along the magnetic field within flux surface) but on the transverse convective cells that represent the resistive interchange/ballooning aspect of SMAPs.

Being a compound quasi-mode, it is understandable that the SMAP phenomenon differs from known isolated eigenmodes in several respects, including an utter lack of clear $\eta$-scaling and the existence of an $\eta$-dependent instability threshold. Thus, our choices of names like ``resistive interchange/ballooning'' are based not on idealized scaling laws but phenomenology.

It remains to be checked with kinetic models whether SMAPs that were encountered in our resistive MHD simulations have any chance of occurring under realistic MCF tokamak plasma conditions. Their main caveat is that they are found only in simulations with sufficiently low thermal conductivity, which ensures that the plasma is able to sustain what in the literature is sometimes called ``thermal misbalance''. At present, we cannot rule out that --- under MCF plasma conditions --- the SMAP phenomenon could turn out to be a mere peculiarity of the MHD model, which does not describe well the effects associated with the free streaming of charged particles along the magnetic field.\footnote{If this is the case, we have at least learned something about the properties of this widely used model. This experience can be useful for setting up simulations and interpreting their results, distinguishing practically meaningful from practically meaningless features. If SMAPs are an unimportant curiosity of MHD, one take-away message would be to avoid running simulations with high resistivity $\eta$ but low viscosity $\nu$ and thermal diffusivity $\chi$. This is one concrete rationale for the commonly used choice $\eta = \nu = \chi$ in MHD simulations.}
However, as discussed in Section~\ref{sec:discussion_exist}, the jury is still out, and we think that it is worth to check whether SMAP-like phenomena can be found in simulations that employ (gyro)kinetic models. There, the resistive destabilization seen in our simulations could be replaced by other mechanisms, such as inverse Landau damping via wave-particle resonances that may feed on, for instance, gradients in the population of thermal or fast ions. For instance, one may investigate whether diamagnetic variants of fishbone oscillations \cite{Coppi86} can exhibit SMAP-like behavior.

Kinetic effects will also be important for investigating practical implications of another interesting observation that was made in our simulations: SMAPs were found to couple to distant rational surfaces (here $q = 2,3,4,5)$, drive magnetic reconnection \cite{Hahm85}, and give rise to sizable magnetic islands, some of which overlapped to produce magnetic chaos. The islands pulsated at the same few-kHz pace as SMAPs, with their X- and O-points appearing at alternating positions in successive pulses. The ``reversibility'' of reconnection on the time scale of these low-frequency modes was facilitated here by a relatively high electric resistivity ($\eta_{\rm e} = 10^{-5}\mu_0 v_{\rm A0} R_0$) and correspondingly strong loop voltage ($-\eta_{\rm e} J_{\rm ref}$ in Ohm's law (\ref{eq:model_cc_efield})) that tends to restore the initial $q$ profile. As discussed in Section~\ref{sec:discussion_nonloc}, similar behavior could be realized in kinetic models via collisionless reconnection, which is truly reversible. Known also to be ``faster'' than resistive reconnection, collisionless reconnection could then also enable higher-frequency modes in the central core to couple to distant rational surfaces near the edge and drive magnetic reconnection there.

Experimental evidence for non-local multi-scale phenomena exhibiting both interchange and tearing parity has been reported experimentally at plasma-frame frequencies in the $\lesssim 20\,{\rm kHz}$ band \cite{Du21}. This is comparable to the frequency $\omega_{\rm A0} \approx 2\pi\times 10\,{\rm kHz}$ of the central plateau of the low-frequency Alfv\'{e}nic continuum branch in our simulations, where we found an electromagnetic counterpart of SMAPs, which we called ``low-frequency compressible Alfv\'{e}nic pulsations (LCAPs)''. Their properties are described in Section~\ref{sec:discussion_summary} and Fig.~\ref{fig:13_flux0.50_Alfvenic-pulse}. In contrast to SMAPs, LCAPs were found in simulations with relatively high thermal diffusivity and low or even zero resistivity. Although they do have a significant compressible component, we suspect that their stronger Alfv\'{e}nicity allows LCAPs to exist even without significant thermal misbalance. Thus, we think that LCAPs are more likely to arise under realistic MCF plasma conditions, and a connection to phenomena like those reported in Ref.~\cite{Du21} seems possible.

\section*{Acknowledgments}\vspace{-0.3cm}

A.B.\ would like to thank Philipp Lauber (Max Planck IPP Garching) for suggesting parameter scans while working on Ref.~\cite{Bierwage26a} that led to the discovery of the unexpected phenomena reported here, Makoto Hirota (Tohoku University) and Zhiyong Qiu (IPP CAS) for motivating us to look deeper into the properties of the model and numerics, and Xiaodi Du (General Atomics) for motivating discussions concerning multi-scale low-frequency modes \cite{Du21}. A.B.\ thankfully acknowledges Yasushi Todo (NIFS) for providing the code {\tt MEGA} \cite{Todo98,Todo05,Todo25}, and Matteo V.\ Falessi (ENEA Fracati) for providing the code {\tt FALCON} \cite{Falessi19b,Falessi20}. Furthermore, the authors thank colleagues at KAIST and KFE for providing the KSTAR tokamak equilibrium and profiles, based on which we constructed our simulation scenarios. Y.-c.G. and G.J.C. are supported by National R\&D Program through the National Research Foundation of Korea (NRF) funded by the Ministry of Science and ICT (grant numbers RS-2022-00155917 and NRF-2021R1A2C2005654). The hybrid simulations were carried out using the supercomputers Plasma Simulator and HPE SGI8600 in the National Institutes for Quantum Science and Technology (QST).

\appendix
\setcounter{figure}{0}
\addtocontents{toc}{\setcounter{tocdepth}{0}}

\section{Additional sensitivity tests}

\begin{figure}
	[tbp!]
	\centering
	\includegraphics[width=0.48\textwidth]{\figures/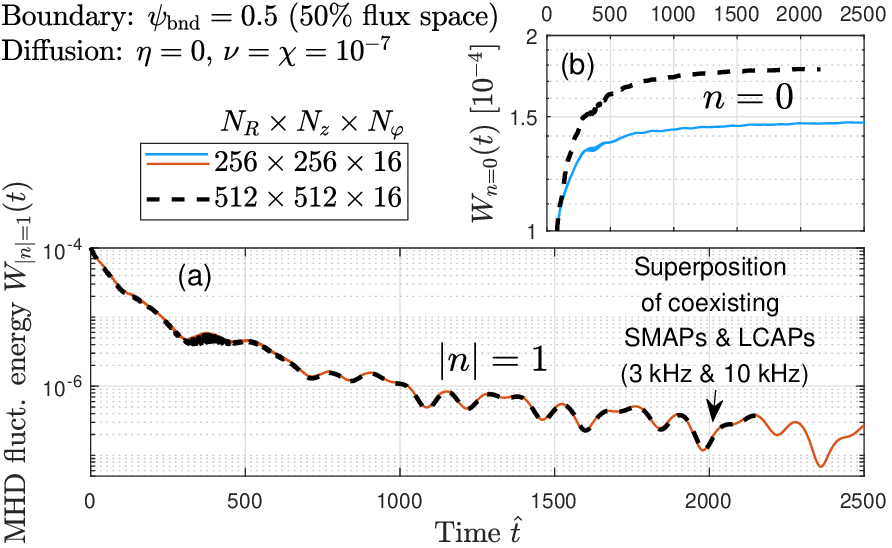}\vspace{-0.25cm}
	\caption{Numerical convergence test with respect to spatial resolution $N_R\times N_z$ for the case with zero resistivity ($\eta=0$) in Fig.~\protect\ref{fig:06_flux0.50cut2_scan-diff}(c).}
	\label{fig:a01_flux0.50_eta0_convergence}%
\end{figure}

\begin{figure*}
	[tbp!]\vspace{-1.3cm}
	\centering
	\includegraphics[width=0.96\textwidth]{\figures/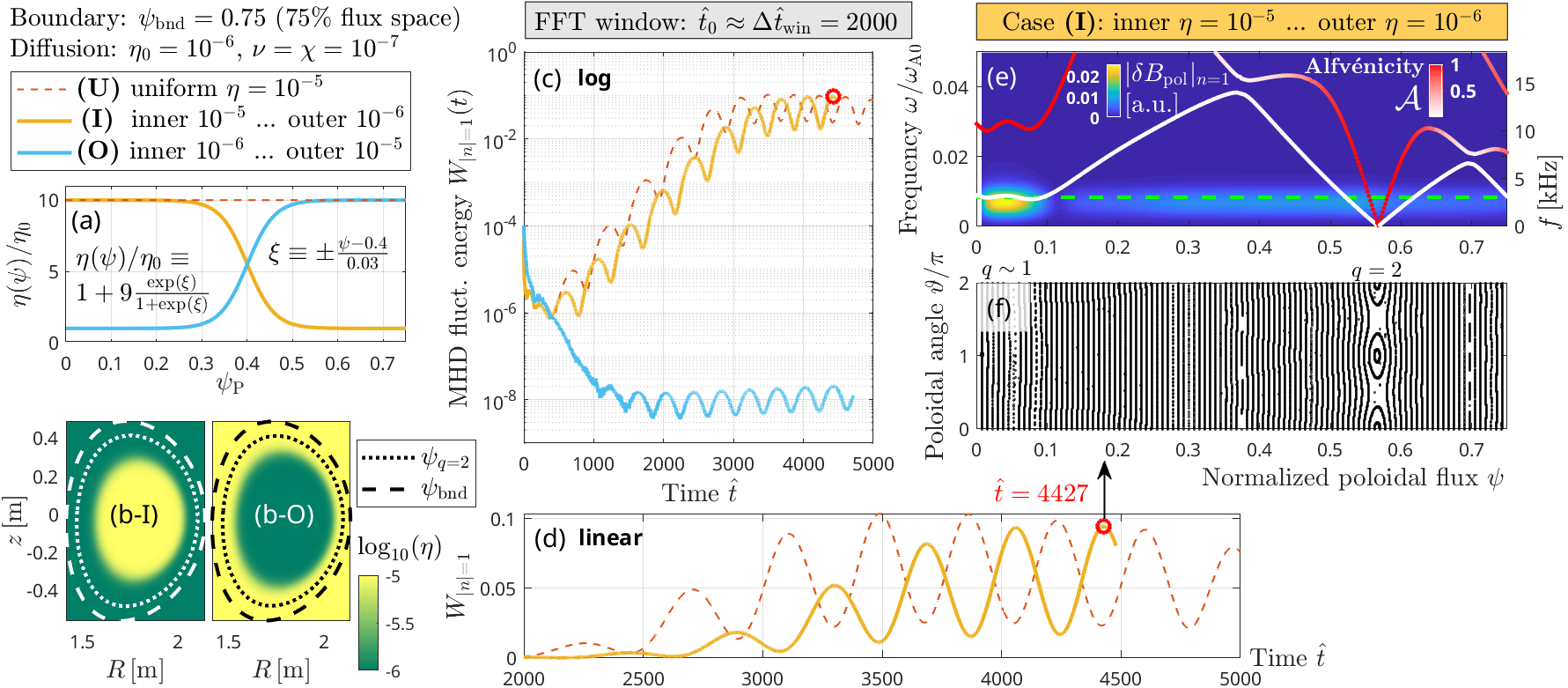}\vspace{-0.25cm}
	\caption{Resistivity-dependence of forced reconnection at the $q=2$ surface driven by core-localized SMAPs in a simulation covering $75\%$ of the flux space as in Fig.~\protect\ref{fig:10_flux0.75cut3_577-evol}. The result from Fig.~\protect\ref{fig:10_flux0.75cut3_577-evol} obtained with uniformly high resistivity $\eta = 10^{-5}$ --- here referred to as Case (U) and plotted using dashed red lines --- is compared with results from simulations with nonuniform resistivity profiles $\eta(\psi)$ as shown in panels (a) and (b). Case (I) (orange) has higher resistivity $10^{-5}$ in the inner region, $\psi < 0.4$, and lower resistivity $10^{-6}$ in the outer region, $\psi > 0.4$. The reverse is true in Case (O) (light blue), where the resistivity is higher in the outer region that includes the $q=2$ surface ($\psi_{q=2}\approx 0.57$). Panels (c) and (d) show the respective time traces of the MHD fluctuation energy $W_{|n|=1}(t)$. For the FFT window $\hat{t} \pm \Delta\hat{t}_{\rm win}/2 = 2000 \pm 1000$, panel (e) shows the spectrogram $|[\delta B_{\rm pol}]_{n=1}|(\omega,\psi)$, which is arranged as Fig.~\protect\ref{fig:10_flux0.75cut3_577-evol}(b). Panel (f) shows the Poincar\'{e} plot for a snapshot of the magnetic field ${\bm B}$ taken at $\hat{t} \approx 4427$, where $W_{|n|=1}$ in panel (d) has a peak (red circle), and the pulsating magnetic island width is expected to be near its maximum (cf.~Fig.~\protect\ref{fig:10_flux0.75cut3_577-evol}(c-4)).}\vspace{-0.3cm}
	\label{fig:a02_flux0.75_eta-nonunif}%
\end{figure*}

\subsection{Convergence test for $\eta = 0$}
\label{apdx:sensitivity_resolution}

Fig.~\ref{fig:a01_flux0.50_eta0_convergence} shows that the coexisting stable slow magnetoacoustic pulsations (SMAP) and stable low-frequency compressible Alfv\'{e}nic pulsations (LCAP) obtained for zero electric resistivity ($\eta = 0$) in Fig.~\ref{fig:06_flux0.50cut2_scan-diff}(c) are accurately reproduced when the spatial resolution in the poloidal $(R,z)$ plane is doubled in both dimensions.

\subsection{$\eta$-dependence of forced reconnection}
\label{apdx:sensitivity_eta-nonunif}

The simulation in Fig.~\ref{fig:10_flux0.75cut3_577-evol} for the medium-sized domain ($\psi_{\rm bnd} = 0.75$) was repeated with a nonuniform resistivity profile $\eta(\psi)$, which is high ($10^{-5}$) either in the inner core (I) or in the outer core (O), and low ($10^{-6}$) elsewhere. The results in Fig.~\ref{fig:a02_flux0.75_eta-nonunif} show the following:
\begin{itemize}
	\item  The core-localized SMAPs ($\psi \lesssim 0.1$) are only weakly affected by the situation in the outer core ($\psi > 0.4$). Consistently with Fig.~\ref{fig:06_flux0.50cut2_scan-diff}(b), they grow rapidly when $\eta(\psi < 0.4) = 10^{-5}$ and are marginally (un)stable when $\eta(\psi < 0.4) = 10^{-6}$.
	\item  Even after several $1000$ Alfv\'{e}n times, the light blue curve in Fig.~\ref{fig:a02_flux0.75_eta-nonunif}(c) shows only the marginal SMAPs. Thus, the $m/n = 2/1$ tearing mode can be assumed to be stable, even in Case (O) with high $\eta(\psi > 0.4) = 10^{-5}$ around the $q=2$ surface.
	\item  Reducing the resistivity around the $q=2$ surface from $10^{-5}$ to $10^{-6}$ as done in Case (I) significantly reduces the width of the $2/1$ magnetic island in Fig.~\ref{fig:a02_flux0.75_eta-nonunif}(f) compared to Fig.~\ref{fig:10_flux0.75cut3_577-evol}(c-4), although it remains clearly visible.
\end{itemize}

\subsection{Role of visco-resistive heating}
\label{apdx:sensitivity_eta-heating}

In the equation of state (\ref{eq:model_cc_pre}) of our self-consistent energy-conserving MHD model, dissipated wave energy is converted into ``heat'', which alters the thermal pressure $P$ (and, with it, the MHD equilibrium). One may thus ask whether visco-resistive heating is an ignorable side-effect or whether it plays a role in the dynamics of SMAPs.

In order to test this, the simulation of the marginally stable case with $\eta = 10^{-6}$ and $\nu = \chi = 10^{-7}$ from Fig.~\ref{fig:06_flux0.50cut2_scan-diff}(c) was repeated while disabling the heating terms of Eq.~(\ref{eq:model_cc_pre}). We found that the first few pulses were unaffected by this modification. Thus, at least unsaturated SMAPs are unaffected by visco-resistive heating.

However, after a few pulses, the fluctuation energy suddenly began to diverge and soon lead to an abnormal termination of the simulation. This behavior is currently not understood. While the said modification of the model violates energy conservation, it was expected to cause merely a ``loss'' of energy without significant consequences for the physical dynamics. This intuition was not confirmed. The role of visco-resistive heating for numerical stability and for the dynamics of quasi-linearly saturated SMAPs remains to be clarified.

\section{Supplementary notes}

This Appendix contains miscellaneous notes that may be relevant for future studies of SMAP- and LCAP-like phenomena in both MHD and kinetic models.

\subsection{Ion acoustic instability (IAI)}
\label{apdx:notes_iai}

In a kinetic model, the core-localized component of our compound mode may become connected to the so-called ion acoustic instability (IAI). For relatively recent work on this subject, see Lesur {\it et al}.~\cite{Lesur14}. Here, we collected some notes of IAI for future consideration.

\begin{figure}
	[tbp!]\vspace{-0.1cm}
	\centering
	\includegraphics[width=0.48\textwidth]{\figures/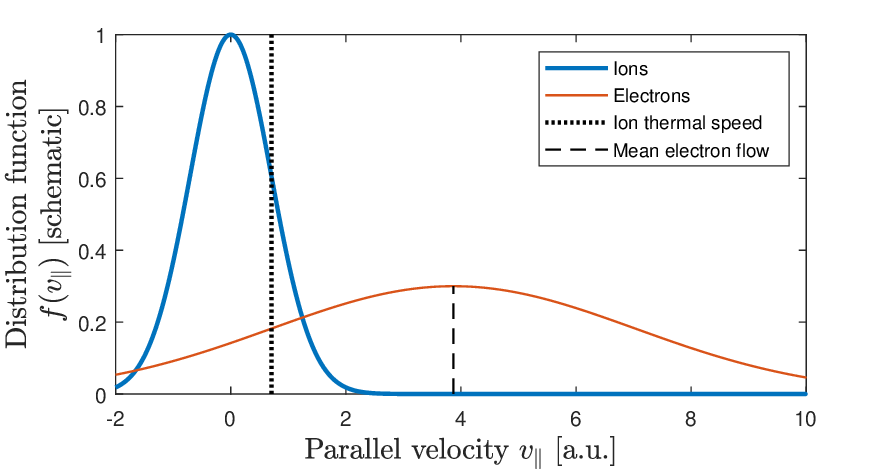}\vspace{-0.25cm}
	\caption{Schematic illustration of the parallel velocity distributions of ions (blue Maxwellian) and electrons (orange shifted Maxwellian) near the threshold for the ordinary ion acoustic instability (IAI). Although we currently see no direct relation to SMAPs studied in the present work, one may looks into a possible connection or interplay between the two phenomena in a kinetic description of the plasma.}\vspace{-0.3cm}
	\label{fig:b01_distrib}%
\end{figure}

IAIs in current-carrying plasmas have been a subject of scrutiny in the 1960s and 70s \cite{Bernstein61, Bernstein64, Wesson73, Hazeltine74}, and continue to be an active area of research in space plasmas (e.g., see Ref.~\cite{LiDion26} and references therein). The instability criterion is schematically illustrated in Fig.~\ref{fig:b01_distrib}, where the electron current $J = -e N_{\rm e}\overline{v}_{\parallel,{\rm e}}$ is so large that the mean electron flow satisfies $\overline{v}_{\parallel,{\rm e}} \gtrsim v_{\rm th,i} + v_{\rm th,e}$, where $v_{{\rm th},\alpha} \equiv \sqrt{2T_\alpha/M_\alpha}$ is the thermal velocity of particle species $\alpha$. This is somewhat reminiscent of a bump-on-tail or two-stream instability, here for two distinct species. Bernstein {\it et al}.\ derived an expression for the critical current density that can be written as (cf.~Eq.~(25) of \cite{Bernstein61} and Eq.~(212) of \cite{Bernstein64}):
\begin{align}
	|J_{\rm crit}| =& 1.97\times e N_{\rm e}\sqrt{T_{\rm e}/M_{\rm e}}
	\label{eq:jcrit}
	\\
	& \times \left[1 + \left(\frac{M_{\rm i}}{M_{\rm e}}\right)^{1/2} \left(\frac{T_{\rm e}/T_{\rm i}}{2.718}\right)^{3/2} {\rm exp}\left(\frac{T_{\rm e}}{2T_{\rm i}}\right)\right]; \nonumber
\end{align}

\noindent which holds with and without strong guide field ${\bm B}$ \cite{Bernstein61}.

If we assume a fully-fledged burning ITER-like DEMO plasma with $T_{\rm e}=T_{\rm i}=30\,{\rm keV}$, number density $N_{\rm e}=10^{20}\,{\rm m}^{-3}$, and effective ion mass $2.5 M_{\rm p}$ (with proton mass $M_{\rm p}$), then the critical current density predicted by Eq.~(\ref{eq:jcrit}) is about $350\,{\rm MA/m}^2$. On a global scale, the plasma seems safe, but the IAI may be expected to arise locally, for instance in narrow current sheets that form in the course of fast reconnection events, in the vicinity of runaway electron beams, and very close to a magnetic axis.

Hazeltine \& Hinton \cite{Hazeltine74} showed that mirror-trapped electrons in magnetically confined toroidal plasmas stabilize the IAI, except within a small region of the minor radius $r$, which they specify as $r/R_0 \ll (M_{\rm e}/M_{\rm i})^{2/3} < 1\%$, where $R_0$ is the major radius of the torus' magnetic axis. Our core-localized slow magnetoacoustic quasi-mode is located around $r/R_0 \sim {\rm few}\,\%$, which is not far from this theoretical threshold.

\subsection{Non-ideal effects}
\label{apdx:notes_diss}

Turning off the visco-resistive heating terms in the equation of state (\ref{eq:model_cc_pre}) does not seem to have a significant influence on SMAPs. This was tested in the marginally stable case with $\eta = 10^{-6}$ and $\nu = \chi = 10^{-7}$. However, the simulation became ill-behaved around $\hat{t} \approx 1500$, after which spurious numerical instabilities arose that grew until the simulation terminated with a {\tt NaN} error.

The KSTAR plasma \cite{Lee23,Lee26} on which our simulation setup is based had $T_{\rm e} \approx 4\,{\rm keV}$, $v_{\rm A0} \approx 4\times 10^6\,{\rm m/s}$ and $R_0 = 1.8\,{\rm m}$. For a Coulomb logarithm $\ln\Lambda \approx 16$, we obtain the Spitzer resistivity and Lundquist number
\begin{subequations}
	\begin{align}
		\eta_{\rm Spitzer} =& \frac{4\sqrt{2\pi}}{3} \frac{Z_{\rm i} e^2 M_{\rm e}^{1/2} \ln\Lambda}{(4\pi\epsilon)^2 T_{\rm e}^{3/2}} \approx 6.5\times 10^{-9}\,\Omega{\rm m}, \\
		S_{\rm Spitzer} =& \frac{\mu_0 v_{\rm A0}R_0}{\eta_{\rm Spitzer}} \approx \frac{1}{7.4\times 10^{-10}} \approx 1.3\times 10^9.
	\end{align}
\end{subequations}

\noindent This is 4 orders of magnitude larger than the value $S \sim 10^5$ at which SMAPs in our simulations reached appreciable growth rates. The standard justification is that the resistivity and other diffusion coefficients in an MHD simulation should be seen as proxies for other dissipative and transport processes that are absent from our simulations but are ubiquitous in real tokamak plasmas, such as microturbulence and high-frequency waves. It would be interesting to clarify whether $S \sim 10^5$ is a realistic measure for the ``effective resistivity'' in the central core of a high-temperature tokamak plasma, specifically in the hybrid scenario with central $q\sim 1$, which are expected to be subject to quasi-interchange and associated flux pumping dynamics \cite{Wesson86, Krebs17} as mentioned in the introduction. Of course, in plasma where resistivity is anomalously enhanced by several orders of magnitude, the same may be assumed to be true for the viscosity and perpendicular thermal diffusivity. That is, as far as SMAPs are concerned, both stabilizing and destabilizing mechanisms are enhanced, so that the outcome is not evident from simple estimates.

\subsection{Mercier stability}
\label{apdx:notes_mercier}

The plasma considered here obviously satisfies the Mercier stability criterion since we have $\beta' < 0$ and $q > 1$ everywhere. Nevertheless, let us quantify the two terms comprising this criterion: magnetic shear stabilization and pressure gradient. The formula for Mercier stability at small inverse aspect ratio ($r/R \ll 1$) is (cf.~Eq.~(6.121) on p.336 of Wesson's book \cite{WessonBook})
\begin{equation}
	\frac{a^2}{8}\left(\frac{q'}{q}\right)^2 + \frac{a^2\mu_0 P'}{r B_{\rm tor}^2}(1 - q^2) = \frac{\overline{s}^2}{8\hat{r}^2} + \frac{{\rm d}\beta}{{\rm d}\hat{r}^2}(1 - q^2) > 0.
\end{equation}

\noindent The minor radius of interest is $\hat{r} \equiv r/a \sim 0.1$ Substituting values for $q \approx 1.05$, $|\overline{s}| \lesssim 0.02$ and $\Delta\beta/\Delta\hat{r}^2 \sim -0.02/0.1$ from Fig.~\ref{fig:01_setup}, the stabilizing shear term is $|\overline{s}|^2/(8\hat{r}^2) \lesssim 0.02^2/(8\times 0.1^2) = 5\times 10^{-3}$, and the pressure gradient term is $\tfrac{{\rm d}\beta}{{\rm d}\hat{r}^2}(1 - q^2) \approx \tfrac{-0.02}{0.1}(1 - 1.05^2) \approx 0.2$, so they differ by a factor of $\sim 40$.

\setlength{\bibsep}{0.6pt}
\addtocontents{toc}{\setcounter{tocdepth}{1}}
\addcontentsline{toc}{section}{References}
\bibliographystyle{unsrt}
\bibliography{references}

\end{document}